May 09, 2021

# Proposal of Bragg Diffraction Imaging on Protein Crystals with Precession-Electron Annular-Bright-Field Microscopy

Tsumoru Shintake

OIST: Okinawa Institute of Science and Technology Graduate University
1919-1, Tancha, Onna-son, Okinawa 904-0495 Japan,
Email: ShintakeLab@oist.jp

In this theoretical study, the author firstly discusses the wave interference of Bragg diffraction inside 3D crystal, followed by quantum mechanical interpretation on the diffraction process, and proves that the interference fringe between Bragg diffraction and the incident beam is identical to the lattice plane. By introducing the beam expander concept, we may explain the image projection mechanism of Bragg diffractions propagating through TEM: Transmission Electron Microscope. In practice, we will take projection images at zone axes, because of its symmetric high-density diffractions associated with flat Ewald sphere and minimum overlap with the neighboring molecules. By precessing the illuminating beam around the zone axis, and introducing annular objective aperture, we select only the kinematic Bragg diffractions which correctly contribute to the real image reconstruction on the 2D electron detector. The outcome should be a positive contrast, clean image with no dynamical diffractions nor Moiré patterns. Importantly, the image should be insensitive to the defocus, and the spatial resolution may not be limited by the spherical and the chromatic aberrations of objective lens. Therefore, the imaging of small molecules, such as, the drug structure will be straightforward. By taking multiple projection images at the different zone axes from many micro-crystals, or from surrounding thin edges of a large crystal, we may reconstruct 3D structure of the larger protein complex in atomic resolution.

## I. Introductions

Phase is missing in Bragg diffractions. This is the central dogma in X-ray/electron crystallography. Without the phase, we cannot reconstruct the structure of the object from the recorded Bragg diffraction pattern. In 1975, Henderson and Unwin [1,2] firstly retrieved phase from 2D crystals (unstained periodic biological specimens) by the Fourier analysis on the noisy real image taken by defocused TEM. They combined with the electron diffraction pattern and determined the structure of purple membrane protein (today known as the light-driven proton pump 'bacteriorhodopsin'). By tilting the specimens, they also determined three-dimensional model of the purple membrane. Those proteins had unique ability to form 2D molecule array.

In protein crystallography, we know most of the proteins tend to form 3D crystals. Ideally protein crystals consist of many identical unit cells, regularly spaced as 3D array, and in the same orientation. By averaging these molecule images, signal-to-noise statistics and contrast will be drastically improved. Therefore, various challenges have been made to directly capture the image of the molecule array by using TEM. For example, Nederlof *et al*. [3] tried to image flash-cooled lysozyme of thickness of 100 nm by 300 kV cryo-EM on a Flacon direct electron detector. While captured images devoid of contrast, Fourier transform showed 3 Å resolution lattice. They noted that the projection images showed Moiré pattern, which manifest themselves in the Fourier transform. Interestingly, period of the Moiré pattern can be longer than the unit cell size. This is serious problem in three-dimensional crystal imaging, whose physical origin should be revealed theoretically and solved in practical instrumentation.

Toots and Skoglund [4] applied the cryo-electron tomography to image lysozyme nanocrystals in the size range of 100 nm using 300 kV TEM. By exploiting crystallographic symmetry inherent in the tomogram, they could achieve resolution in the range of 15Å.

Until today, there are no reports of successful structure determination of 3D protein crystals via electron imaging at the spatial resolution comparable to the diffraction methods in the range of a few Å.

There are intrinsic problems behind of these outcomes, as follows.

(a) The atomic density of proteins is close to the water, and thus the image contrast is very low in real space imaging.

(b) Protein crystals typically contain 30-75% solvent (water) by volume. Those water molecules randomly scatter the electrons, and thus cause noisy background.

(c) The dynamical electron diffraction causes various image confusions, such as negative contrast.

(d) Ewald sphere is almost flat due to extremely short wavelength of 200~300 kV electron beam, and thus the Bragg diffraction is highly sensitive to the incident beam angle. However, due to the sample damage there is not much time to precisely tune the crystal angle in practical experiments.

(e) Real image can be easily deteriorated by the sample vibration or drift.

(f) Moiré patterns were frequently observed, and thus image processing became difficult.

(g) Protein microcrystals are not perfect, contains mosaic structure and disorders.

To overcome above difficulties, the author proposes the following strategies.

(i) We focus on taking projection images at zone axes of low index; [100], [010], [001], [110], [011], [101], *etc*. As discussed in Appendix A1, there are 13 unique axes. Since wavelength of the electron is very short, Ewald sphere is flat (a very large radius), so that electron diffraction spots generally "light up" at zone axes. And overlapping between neighboring protein molecules becomes smallest at zone axes, and thus two-dimensional array becomes visible. This is extremely advantageous in projection imaging, while we must handle the harmful dynamical diffractions, as follows.

(ii) We introduce beam precession in the upstream illuminating system just same as PED (Precession Electron Diffraction) method, which reduces dynamical diffractions. We take projection image under precession beam. The precession motion does not deteriorate the crystal image if we do not use downstream de-scan coils (the image shift coils).

(iii) We introduce an annular objective aperture (filter) to select kinematic Bragg diffractions and eliminate various noises from the sample; remaining dynamical diffractions, scattering by disordered unit cells at the surrounding surface of microcrystal, diffractions from the mosaic structures in wrong orientations, random diffraction from small ice-crystals and the structure water, and high frequency diffractions from the higher order Laue zones (HOLZ) which cause Moiré patterns.

(iv) We take single image per micro-crystal at high electron dose (3~10 e/Å2) to maximize signal-to-noise statistics. We need multiple microcrystals to collect enough images in different zone axes. R&D for efficient angle tuning will be the key for success.

(v) Alternatively, we may collect multiple projection images at surrounding thin edges of a larger crystal, by rotating crystal to successive zone axes.

In this paper, we treat only the primitive cubic cell; $a = b = c,\ \alpha = \beta = \gamma = 90^\circ$, while all discussions should work for different crystal system, for example $a \neq b \neq c$, $\alpha \neq \beta \neq \gamma \neq 90^\circ$. If the crystal structure has the translational symmetry, Fourier transform works, and thus we can discuss diffractions in reciprocal space (Fourier space) in the same manner. Each unit cell generally consists of multiple molecules in certain space-group symmetry, which we also hope to observe directly under TEM by the current imaging method. Regarding space-group symmetry, Dautera and Jaskolski published well written introduction [5].

## II. Learning from Micro-ED

Recently Micro-ED (Microcrystal electron diffraction) technique has become available, and many protein structures have been determined [6,7,8]. It is developing into a powerful alternative method to X-ray crystallography on single biomolecules. While Micro-ED has successfully used direct method to process data (diffraction data to 3D structure reconstruction), currently limited to small molecules. This is due to missing phase in recorded Bragg diffractions. X-ray crystallography overcomes this using Isomorphous Replacement (IR: heavy metal soaking of crystals) and anomalous scattering (AS: X-ray wavelength is altered to resonate on heavy metals which shift phases, from which we retrieve the original phases), but MicroED has not implemented those phasing methods so far, an area of research that requires further investigation [6]. We challenge this problem in this paper through direct imaging on the array of molecules.

Important information from Micro-ED community is on the damage threshold. The protein microcrystals hold their detail structure up to dose 3~9 e/Å$^2$ by electron beam of 200 kV energy at cryo-cooled condition [6]. The optimum thickness of the protein crystal is 100 ~ 500 nm range.

## III. Interference fringe created by Bragg diffraction

### Interference Fringe and Lattice Plane

Bragg diffraction was first discovered by Lawrence Bragg and his father William Henry Bragg in 1913 [9]. They reported "*attempt to use cleavage planes as mirrors, and it has been found that mica gives a reflected pencil from its cleavage planes as mirrors*." This is important message; the Bragg diffraction is "reflection phenomena" from regularly arranged atoms of the crystal. From the careful experiments, they found famous formula:

$$2d_a \sin\theta = n\lambda \qquad (1)$$

where λ is X-ray wavelength, *n* is integer, $d_a$ is atom spacing of crystal. $\theta$ is the reflection angle (= incident angle). This finding was the starting point of long successful history of the crystallography, first in minerals and later in protein crystallography.

Here we perform a thought experiment using Micro-ED setup. Figure 1 shows schematic illustration of microscopic view around the microcrystal, where the lattice plane (Bragg plane) of (*h k l*) reflects the incident wave into the reflection angle of $2\theta$. Set of integers (*h*, *k*, and *l*) are the Miller indices of the lattice plane. The illustration plane (sheet of the paper of Fig. 1) is chosen as normal to the lattice plane, and the incident wave and Bragg diffraction are in the same plane. The crystal is aligned as zone axis in z-direction. The incident waves are coherent and in the same phase.

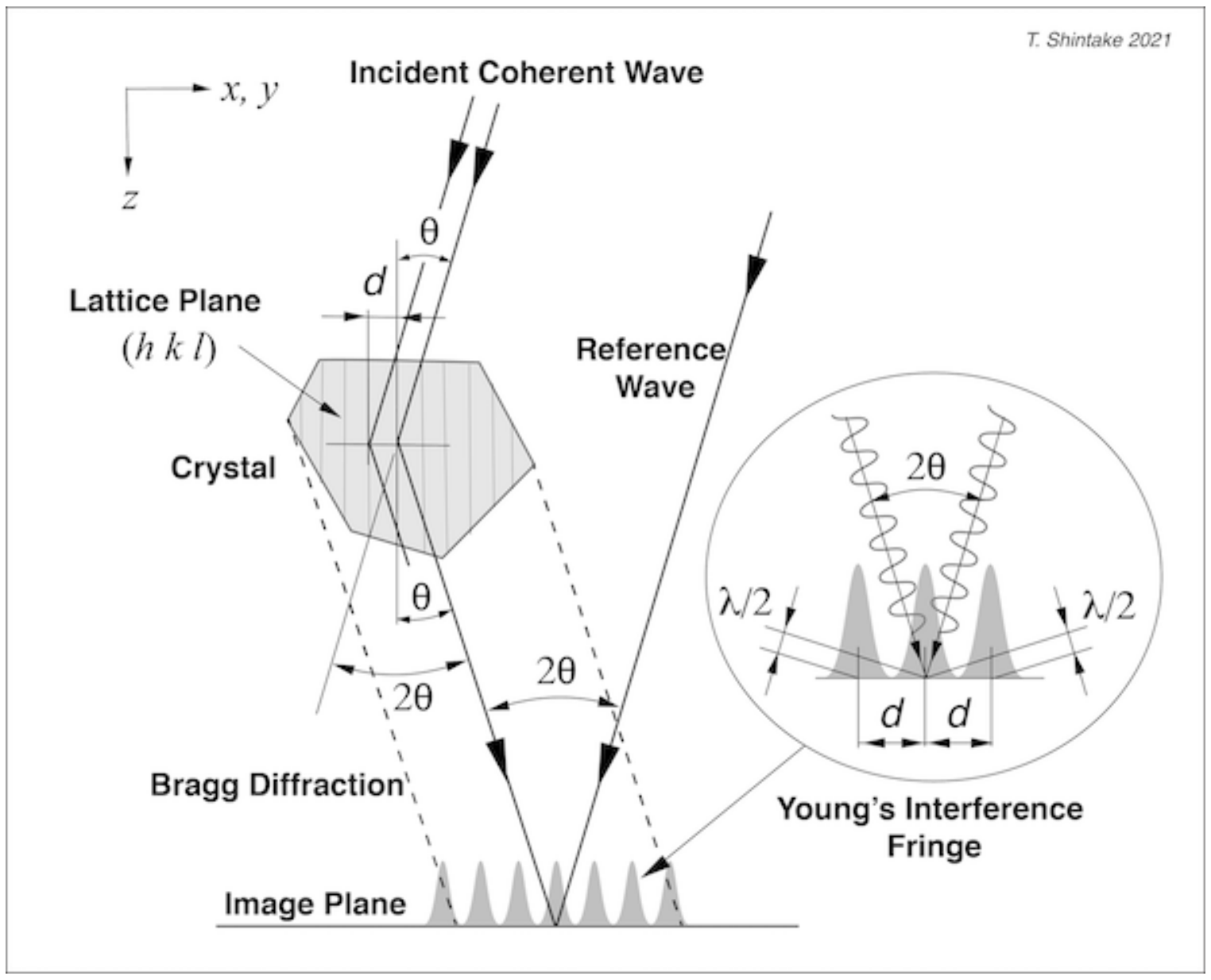

Fig. 1. Bragg diffraction from a lattice plane of (*h k l*), which overlaps with the reference wave and creates the interference fringe at the image plane located downstream very close to the crystal.

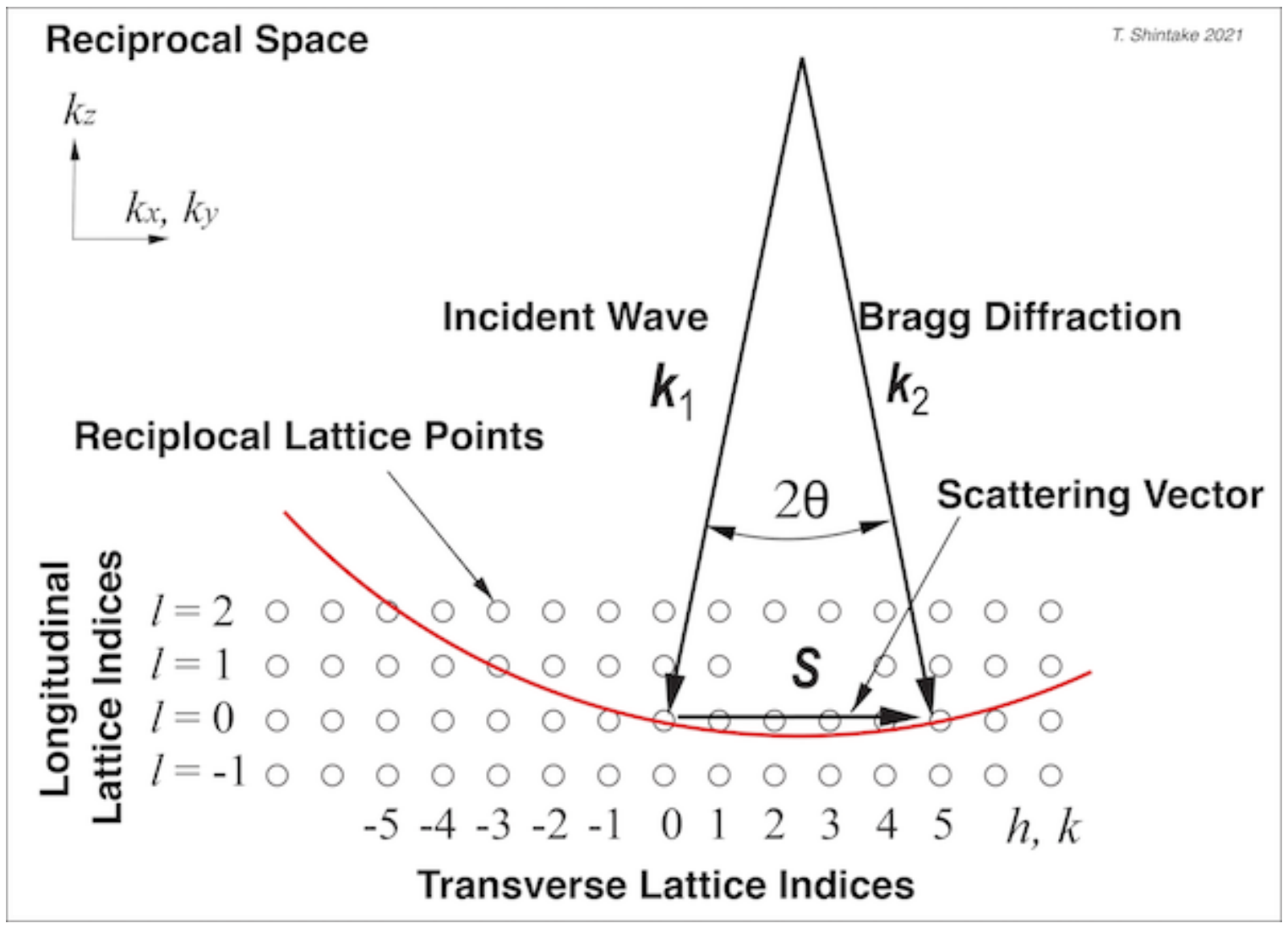

Fig. 2. Wave vectors on the reciprocal space. Red curve represents Ewald sphere, the curvature is emphasized in this drawing. In electron crystallography, Ewald sphere becomes a flat plane because of extremely short wavelength.

Fig. 2 shows the wave vectors in the reciprocal space. The incident wave vector $\mathbf{k}_1$ is scatted into $\mathbf{k}_2$ by the reciprocal lattice point (*h k l*). The scattering vector is given by

$$\mathbf{S} = \mathbf{k}_2 - \mathbf{k}_1 \tag{2a}$$

$$|S| = \frac{2\pi}{d} = 2\,k_0\,sin\theta \tag{2b}$$

We assumed there is no energy gain or loss in the diffraction process, and thus $\mathbf{k}_1$ and $\mathbf{k}_2$ have the same vector length $k_0$.

From Eq. (2b), the lattice spacing becomes,

$$d = \frac{\lambda}{2\,sin\theta} \qquad (3)$$

We must note that $d$ is always smaller than the unit cell size (see Fig. 4).

We place an image plane very close to the crystal as shown in Fig.1. Bragg diffraction and reference wave (incident coherent electron beam) will overlap and create interference fringe on the imaging plane just same as Young's interference fringe (see Appendix 3).

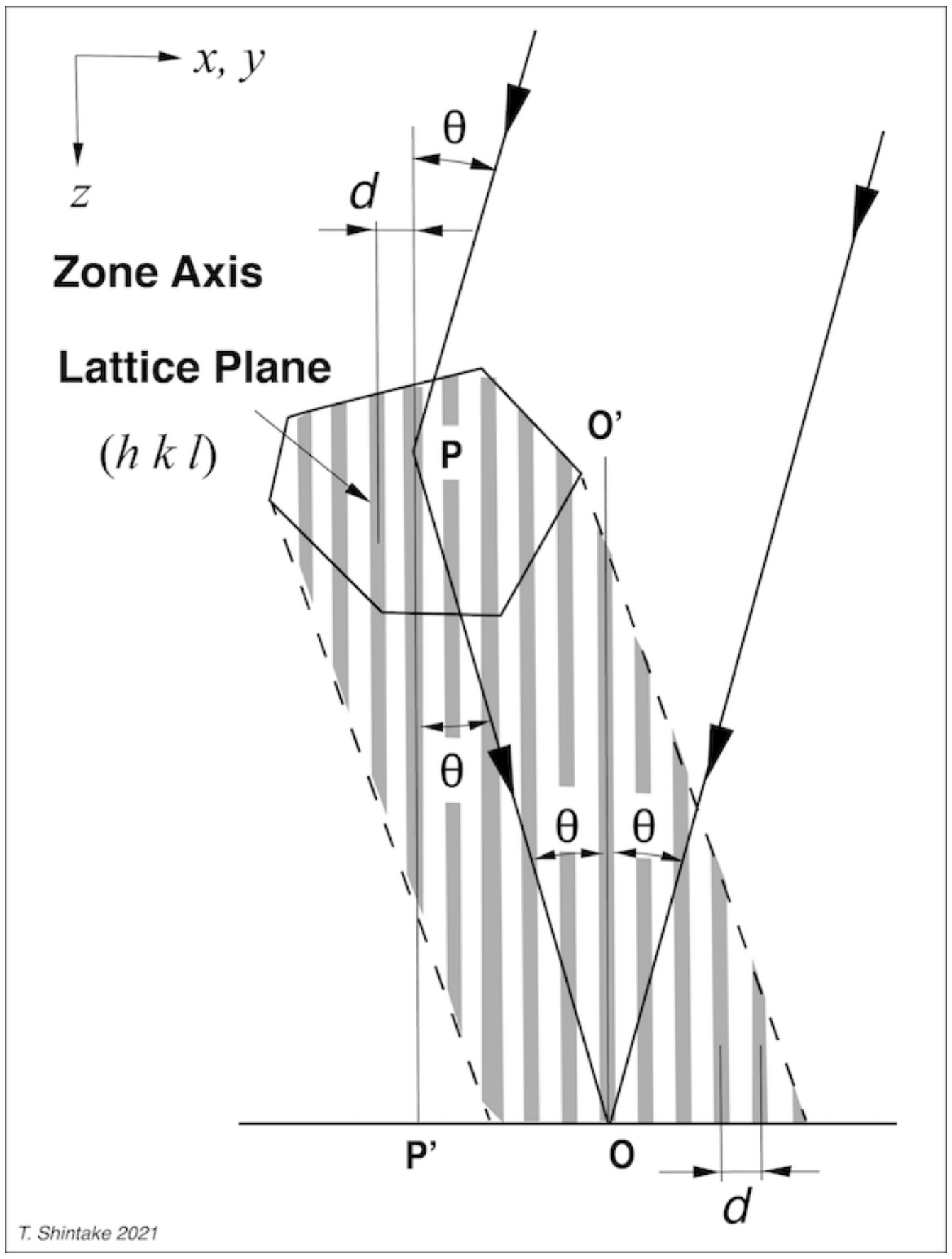

Fig. 3. The extended interference fringe matches with the lattice plane.

As shown in Fig. 3, when we move the image plane upward, we observe the same interference fringes with the same pitch, because the opening angle between two waves stays constant. On the mean plane OO' (mean angle of two waves), the path length difference of two waves stays same, thus bright zone of center fringe stays always on the mean plane OO'. The lattice plane PP' and mean plane OO' are parallel. We continue to move the image plane into the crystal, we should still observe the interference fringe, because there exists diffracted wave from the upstream atoms and incident wave running through the crystal and they will interfere. When the image plane reaches to the reflection point $P$, the path difference between two waves becomes zero. Therefore, we may conclude that the lattice planes must perfectly overlap with the bright zone of the interference fringes inside the crystal. The interference fringe is identical to the crystal lattice.

This discussion is closely related to the volume hologram, first treated by Kogelnik [10] in 1969 in the coupled-wave theory.

## Quantum Mechanical Interpretation

In quantum mechanics, a running electron is written by the de Broglie wave (the electron wave, the matter wave, or the probability wave), whose wave vector is $k = 2\pi/\lambda$, the wavelength is $\lambda = h/p = h/\gamma m_e v$, where $\gamma m_e$ is relativistic mass of the electron. In our crystal experiment, the electron wave $\Psi_1$ incidents and scatters into $\Psi_2$.

$$\Psi_1 = A_1\ e^{i(\mathbf{k}_1\cdot\mathbf{r}-\omega t)} \tag{4a}$$
$$\Psi_2 = A_2\ e^{i(\mathbf{k}_2\cdot\mathbf{r}-\omega t)} \tag{4b}$$

According to Born rule [11], the expectation value of the observable array of molecule, or the scattering probability can be written as follows.

$$\langle n\rangle = \langle\Psi|\rho(\mathbf{r})|\Psi\rangle \tag{5a}$$
$$\Psi = \Psi_1 + \Psi_2 \tag{5b}$$

where $\rho(\mathbf{r})$ is the density of scatterer, i.e., atoms in molecular array. This is the quantum mechanical interpretation of Bragg diffraction phenomena.

From Ea. (5), the scattering intensity becomes,

$$\begin{aligned} I &= \int \rho(\mathbf{r})\,\Psi\cdot\Psi^*\,dv \\ &= \int \rho(\mathbf{r})\left(A_1^2 + A_1A_2\,e^{-i\mathbf{S}\cdot\mathbf{r}} + A_2A_1\,e^{+i\mathbf{S}\cdot\mathbf{r}}\right)dv \\ &= I_0\int \rho(\mathbf{r})\left(1 + \frac{A_2}{A_1}e^{-i\mathbf{S}\cdot\mathbf{r}} + \frac{A_2}{A_1}e^{+i\mathbf{S}\cdot\mathbf{r}}\right)dv \end{aligned} \tag{6}$$

where intensity of the incident beam is normalized as $I_0 = A_1^2$, and we assume $A_2 \ll A_1$. The first term represents the central beam in the diffraction pattern. Second and third are the scattering amplitudes of the form factors identical to $\mathbf{K}(\mathbf{S})$, $\mathbf{K}(-\mathbf{S})$ in Eq. (10). These two form factors have the same amplitude, and they appear one by one according to crystal orientation in the diffraction experiment [12].

The probability to find an electron is given by $|\Psi|^2 = A_1^2 + 2A_1A_2\,cos\phi + A_2^2$, which gives periodical amplitude modulation as shown in the vertical bright-dark zones in Fig. 4. The fringe pattern (lattice) periodicity is three times higher than the unit cell size in Fig. 4, and thus Miller indices becomes (300). At the bright zones, there is more chance to find electrons, which will be scattered by the molecules and exchange the momentum through Coulomb force. At dark zones, the scattering chance is low. Therefore, we may predict that the bright peak should come to the center of molecule density, and the phase of interference fringe of $|\Psi|^2$ should meet with the structure phase. In other words, the diffraction wave chooses its phase to match $|\Psi|^2$ with the lattice plane. This is intrinsically included in Eq. (6) as Fourier transform of the molecule density.

In this coherent scattering, there no energy exchange. If the incident electron couples to the orbital electron of the atoms, it may excite orbital electron to higher state, accordingly the incident electron loses its energy. This is inelastic scattering, which causes background noise, and does not contribute to the image reconstruction.

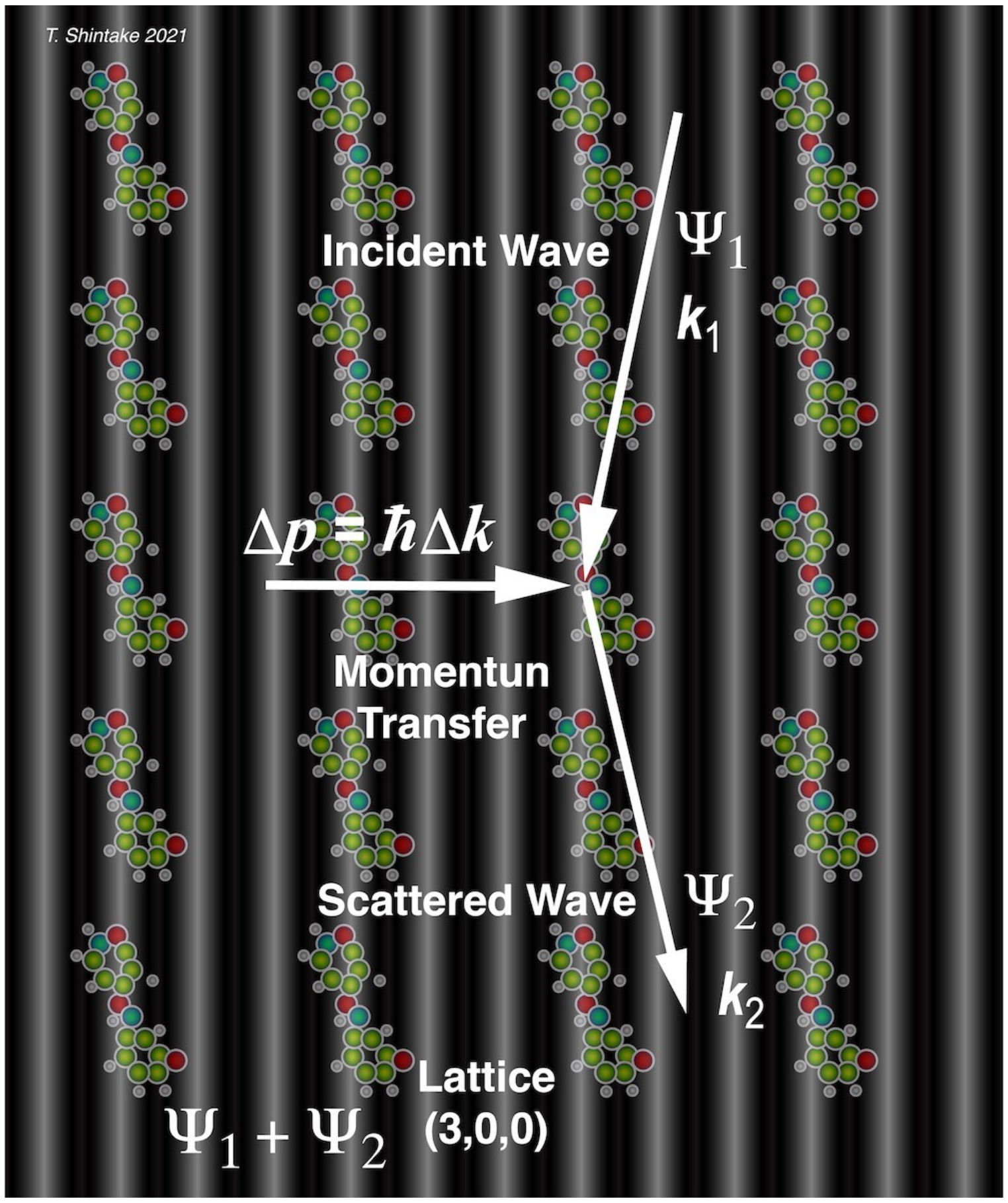

Fig. 4. White vertical zones represent the interference fringe of overlapping two waves: $\Psi_1, \Psi_2$. Scattering probability takes peak at Bragg condition, where the momentum will be transferred from molecules to the electron wave.

## IV. Magnification of Interference Fringe through Beam Expander

In the previous section, we found the interference fringe of Bragg diffraction is identical to the lattice plane. If we magnify the interference fringe using todays advanced electron microscope of atomic resolution, we will be able to observe the crystal lattice. We must note that we are not trying to obtain image of the single atom (or molecule) one by one in the crystal but will observe image of the lattice (section image of lattice plane).

Here we discuss a thought experiment using the beam expander concept (refer to Appendix A2). The lens system is exactly same as the common TEM, and it works as beam expander when the parallel beam is injected.

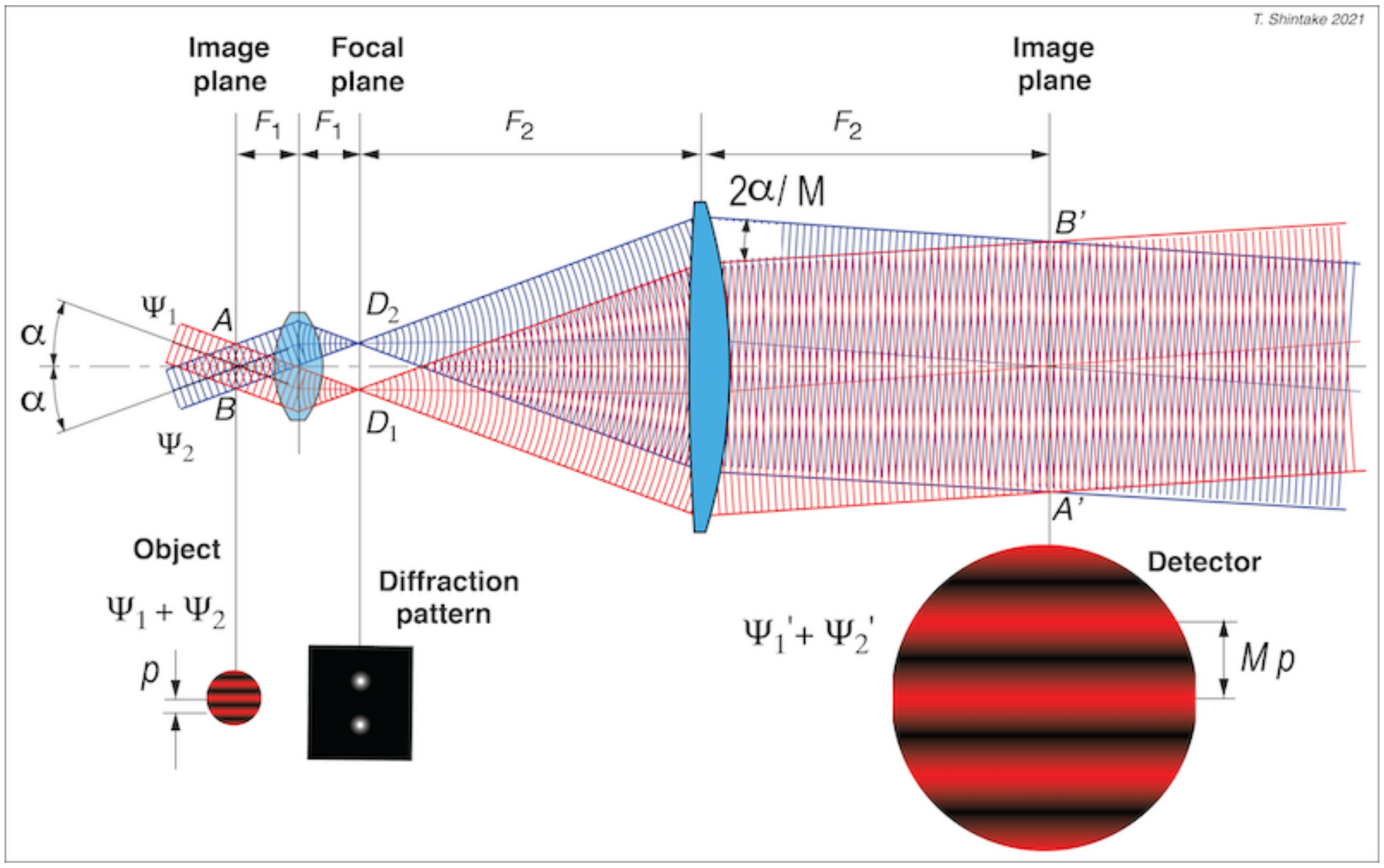

Fig. 5. Two coherent beams are expanded through the beam expander, at the same time the interference fringe is magnified.

Now we try to transfer two coherent beams $\Psi_1, \Psi_2$ through the beam expander. When we inject two beams at the object plane with opening angle 2α, the interference fringe pitch becomes

$$p = \frac{\lambda}{sin(2\alpha)} \sim \frac{\lambda}{2\alpha} \tag{7}$$

In case of the electron microscopes, the angle of beam is always very small, i.e., $\alpha = 0.001 \sim 0.01$ rad, thus approximation $sin(2\alpha) \sim 2\alpha$ works well. Two beams will be expanded to $\Psi_1', \Psi_2'$ and overlap again at downstream image plane, the fringe pitch becomes

$$p' = \frac{\lambda}{sin(2\alpha/M)} \sim M\,\frac{\lambda}{2\alpha} = M \cdot p \tag{8}$$

We must note that the crossing angle becomes very small and thus fringe pitch becomes large. If we place two-dimensional detector at image plane, we will observe the magnified interference fringe as shown in the circle in Fig. 5.

Importantly, at the focal plane, we have two bright spots: D1, D2, corresponding to the Bragg diffraction spot and the central beam. As discussed in Appendix A2, object A-point is projected to A'-point at the image plane, and B to B'. The length of the paths through A-D1-A' and A-D2-A' are same. In the same way, B-D1-B' and B-D2-B' are same. Therefore, there is no phase difference between two waves $\Psi_1, \Psi_2$ from the image plane to the image plane. The interference fringe at the image plane (object) is reconstructed in right phase at the image plane (detector) through this reason.

Interestingly, even if we move z-location of the detector around the image plane, the recorded interference fringe should be unchanged. The image is insensitive to the defocus, importantly there is no CTF function in imaging of Bragg diffractions.

As shown in Fig. 6, we place a crystal at the object and illuminate by the coherent electron wave $\Psi_1$ from angle of θ, which is reflected by lattice (*h k l*) into angel of θ as the diffracted wave $\Psi_2$. Sine we decided to take projection image at zone axes, the lattice is aligned to the microscope axis. Two waves propagate through the lens system and create interference fringe on the detector which is replica of the lattice (*h k l*).

In the famous Tonomura experiment [13], illuminating a double slit (bi-prism) with coherent electron beam at low current, and the electrons were detected on two-dimensional detector. We may believe individual electron run through the slit one-by-one, while we do not know which side of the slit. However, after enough accumulation time, a large number of electron events created interference fringe. This experiment provided a simple but confident proof of the wave-particle duality. When an electron is running in free space, the electron wave $\Psi$ (probability wave function) is propagating, and it can go through both sides of the slit at the same time and create interference fringe. When the wave incidents to the detector, the wave function collapses into one spot as "an event", depositing energy into the silicon detector. The event probability is given by $|\Psi|^2$.

At the focal plane of Fig. 6, we have two spots, i.e., the incident wave $\Psi_1$ (0-th beam or central beam) and Bragg diffraction $\Psi_2$. The phase information is preserved in these spots, thus create interference fringe on the detector at correct phase (bright zone location). This gives a hint of "annular objective aperture" placed at the focal plane, which allows to pass $\Psi_1$ and $\Psi_2$, and reject unwanted noise: dynamical diffractions and others.

We must note that the central beam appears at lower side from the axis. In the precession electron diffraction (PED), the diffraction image is collected by tilting the incident beam around the axis. Optics is tuned by increasing lens power, i.e., shortening $F_2$, the diffraction pattern is imaged on the detector. To keep the central beam on the axis, the diffraction image position is shifted by using downstream corrector coil (the image shift coil). However, in our imaging method the image shift scheme will break the lattice image on the detector. We keep the imaging optics unchanged during the beam precession. We do not use image shift coil.

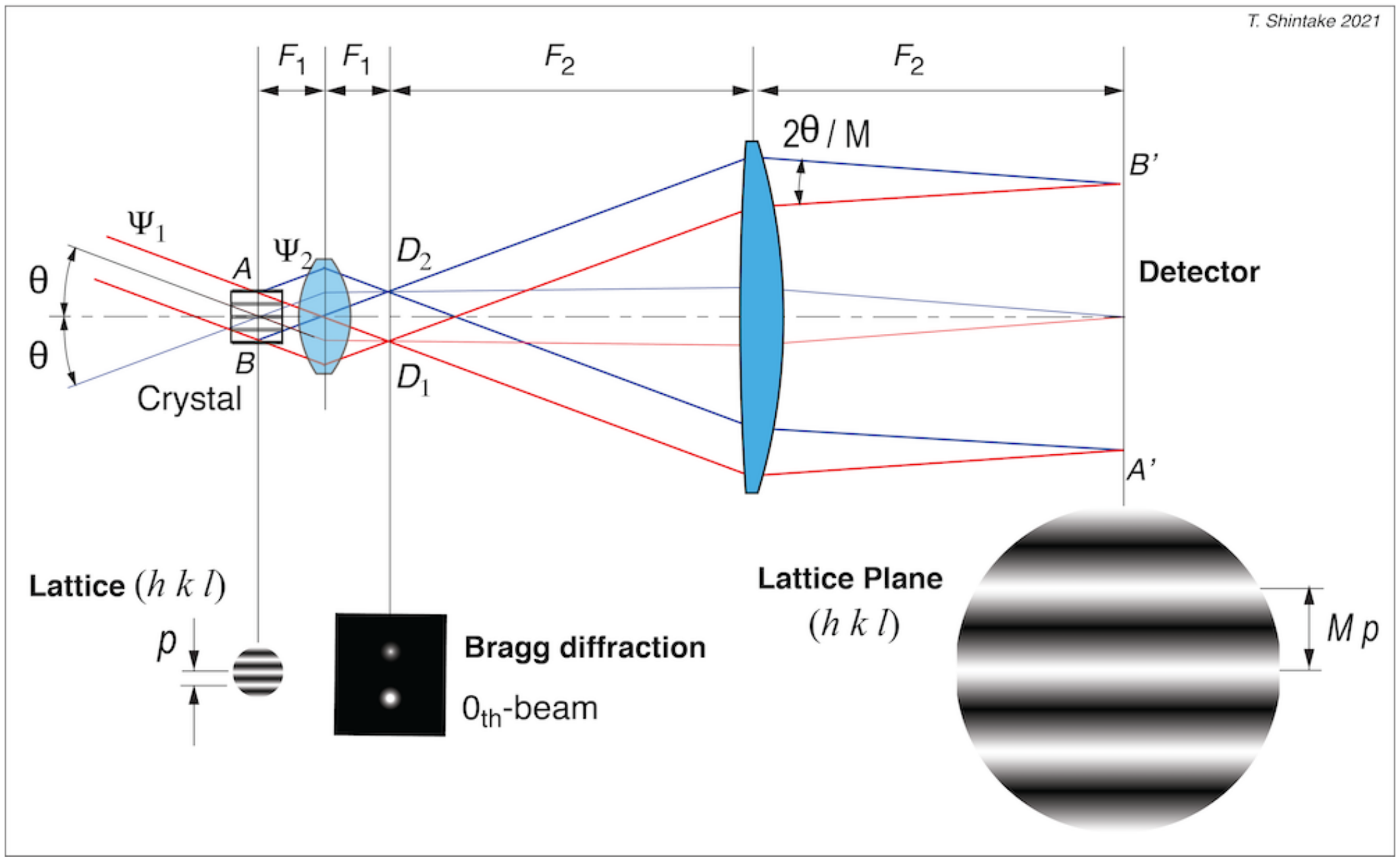

Fig. 6. We place a crystal in zone-axis alignment and illuminate by the coherent electron wave $\Psi_1$ from angle of θ, which is reflected by lattice (*h k l*) into angel of θ as the diffracted wave $\Psi_2$. Through the lens system, we have magnified image of the lattice plane on the detector.

## V. Scattering by Crystal and Molecule Form Factor

The proposed imaging method in this paper does not rely on the diffraction analysis. However, to understand what we observe in the crystal under electron microscope, it is very useful to follow the theoretical basis of the X-ray crystallography [12]. For simplicity, the normalization factors in Fourier transforms are dropped in the following discussions.

The crystal is made of translational copy of unit cell. There are various crystal systems and space groups in protein crystals. As long as the crystal has translational symmetry the following discussion should work universally. The total density distribution can be written as

$$\rho_{crystal}(\mathbf{r}) = \sum_{u=0}^{n_1}\sum_{v=0}^{n_2}\sum_{w=0}^{n_3}\rho(\mathbf{r}-u\mathbf{a}-v\mathbf{b}-w\mathbf{c}), \tag{9}$$

where we suppose the crystal has translation vectors **a**, **b** and **c** and contains number of unit cells: $n_1$ in the **a** direction, $n_2$ in the **b** direction, and $n_3$ in the **c** direction. In the X-ray crystallography, $\rho(\mathbf{r})$ is the electron density in the unit cell. Only the electron contributes to X-ray Thomson scattering, and atoms do not participate due to extremely heavy mass. In electron crystallography, $\rho(\mathbf{r})$ is the density of scattering kernel associated with Coulomb potential created by positive nuclei and negative electron shield. We may approximate $\rho(\mathbf{r})$ as the atomic density distribution in the molecule.

The scattering amplitude from the crystal is given by

$$\mathbf{K}(\mathbf{S}) = \int_v e^{i\,\mathbf{S}\cdot\mathbf{r}} \rho_{crystal}(\mathbf{r}) \cdot d\mathbf{r} \tag{10}$$

where **S** is the scattering vector: $\mathbf{S} = \mathbf{k}_2 - \mathbf{k}_1$. By the convolution theorem,

$$\mathbf{K}(\mathbf{S}) = \mathbf{F}(\mathbf{S}) \times \sum_{u=0}^{n_1} e^{i\,u\mathbf{a}\cdot\mathbf{S}} \times \sum_{v=0}^{n_2} e^{i\,v\mathbf{b}\cdot\mathbf{S}} \times \sum_{w=0}^{n_3} e^{i\,w\mathbf{c}\cdot\mathbf{S}} \tag{11}$$

where $\mathbf{F}(\mathbf{S})$ is the molecule form factor of unit cell. For N-atoms system (a molecule), it can be written as follows.

$$\mathbf{F}(\mathbf{S}) = \sum_{i=1}^{N} f_i \exp(i\mathbf{S} \cdot \mathbf{r_i})$$

$$= \int_{cell} \rho(\mathbf{r}) \cdot \exp(i\mathbf{S} \cdot \mathbf{r})\, dv \tag{12}$$

Mathematically, Eq. (12) is equivalent to the 3D Fourier transform. If we know the molecule form factor $\mathbf{F}(\mathbf{S})$, we can deduce the molecule structure $\rho(\mathbf{r})$ by the reverse Fourier transform. We must note that $\mathbf{F}(\mathbf{S})$ is complex number, i.e., must include the phase; $\mathbf{F} = |\mathrm{F}| \cdot \exp[i\alpha]$.

The three summations in Eq. (11) represent the interference effect, which acts as the sampling function on the form factor, i.e., the lattice points in Fig. 2. In case of Micro-ED, the wavelength of electron wave becomes much short, thus the Ewald sphere become very flat as shown in Fig. 7, where the unit cell size 10 nm cubic, the transverse crystal size 500 nm square, and 200 nm thickness and the electron beam energy 200 kV are assumed. The lattice point has finite size of $\Delta_h = 1/n_x = 0.02$, $\Delta_l = 1/n_z = 0.05$. If the incident beam angle is correctly aligned to the crystal axis, Ewald sphere may match to the lattice points around $h$ = –20 to 20 because of the finite longitudinal size of lattice point: $\Delta_l$, and thus the available resolution will be around 0.5~1 nm. To collect diffractions at higher resolution, we need to tilt the incident beam angle.

As discussed in the next section, to reach 1 Å resolution, we must collect lattice index up to 100, where the required tilt angle should be $\theta = \pm\,12.5$ mrad (see Fig. 8). Ewald sphere has disk shape with large radius, like an umbrella almost a flat disk, and we tilt the umbrella handle along circle trajectory, i.e., as the precession motion.

In the case of 2D crystal, the lattice points are longitudinally elongated rod and thus Ewald sphere meets all transverse lattices at once. Therefore, it is possible to achieve high resolution without tilting the beam.

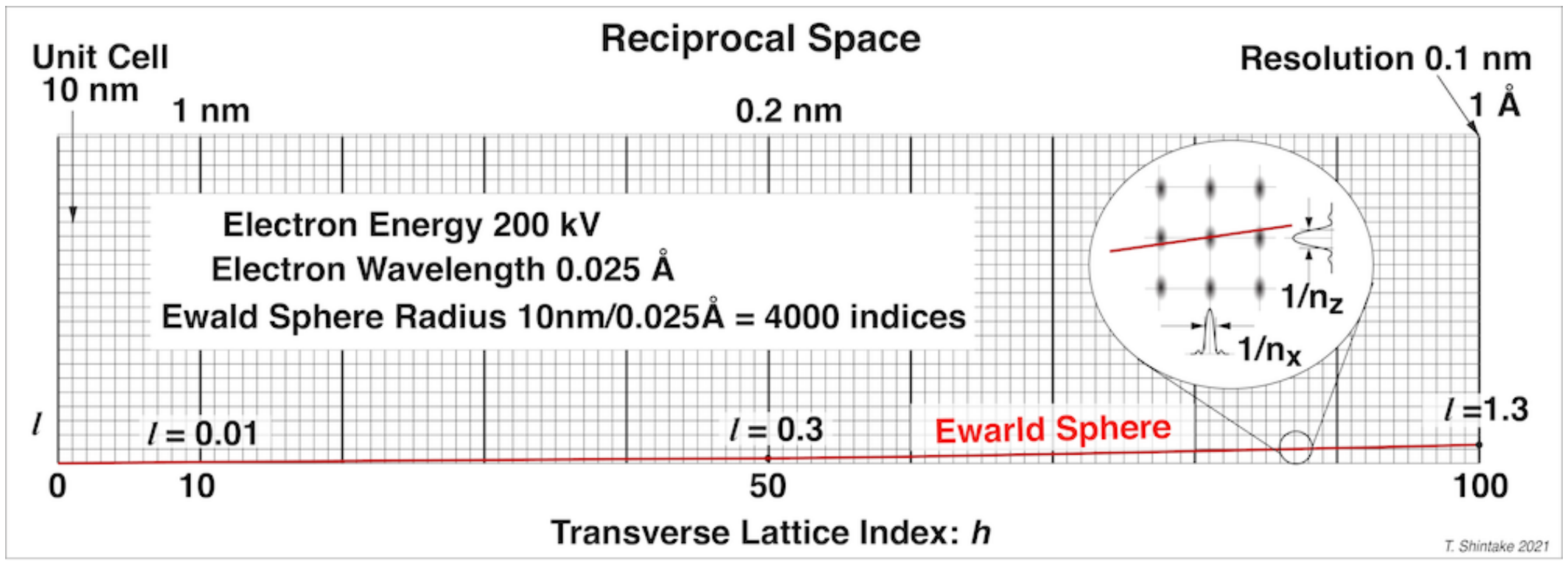

Fig. 7 Reciprocal space (right half) and Ewald sphere for 10 nm unit cell with 200 kV electron beam.

Applying precession beam, we collect the lattice points normal to the incident beam, i.e., mostly $l = 0$. From Eq. (12), we have

$$\mathbf{F}(h,k,0) = \int_{xyz} \rho(x,y,z)\,dz \cdot \exp[i2\pi(hx+ky]\,dxdy$$

$$= \int_{xy} \rho(x,y) \cdot \exp[i2\pi(hx+ky]\,dxdy \qquad (13)$$

where $\rho(x,y)$ is two-dimensional projected density. With inverse Fourier transform, we may find the projected density,

$$\rho(x,y) = \sum_h \sum_k \mathbf{F}(h,k,0) \cdot \exp[-2\pi i(hx+ky)] \qquad (14)$$

In our proposed imaging method, the summation of Eq. (14) is automatically performed by continuous image acquisition of interference fringes during the beam precession.

## VI. Precession Electron Diffraction and Annular Objective Aperture

It is well known that when the electron beam is injected to a mineral crystal along its zone axis, the dynamical diffractions dominate. Physical origin is the multiple diffractions by lattices around zone axis, creating traveling wave (Bloch wave) inside the channels between atomic arrays and comes out as the exiting beam at the downstream side of the crystal [14]. The exiting beam is two-dimensional array of bright spots (approximately represents projection image of crystal), where each dot emits spherical waves in the same phase (they are all coherent and same phase). They interfere each other, and create strong beam in multiple directions, i.e., Bragg diffractions from 2D crystal like. When we observe crystal under TEM bring the focus near to the crystal surface, we may observe array of dots, and we might recognize as “atomic array of the crystal”.

In case of protein crystals, the unit cell size is much bigger, where a large number of atoms are contained in each unit cells. The above-mentioned exiting beam does not provide clear projection image, because Bloch wave consists of multiple plane waves of different propagating angle and thus multiple projected protein images will overlap, and the exiting beam causes image confusion. We must eliminate the dynamical diffractions.

The precession electron diffraction (PED) method has been used to eliminate the dynamical diffractions, and thus we obtain the kinematic diffraction patterns, which we can utilize crystal structure analysis. PED was firstly introduced by Vicent *et al*. [15], where the incident beam was titled and rotated around the axis. Focal plane image (diffraction pattern) was collected with de-scan coils (the image shift coils) at downstream which bring the central beam back to the center, as a result they could obtain correct kinematic diffraction pattern. Oleynikov *et.al.* performed precise numerical simulations and demonstrated PED reducing dynamical diffractions efficiently [16].

As we learned in Fig.3 and Fig.4, the projected lattice image does not cause position shift associated with change of the incident beam angle. During the precessing beam, series of Bragg conditions matches and light up, and those kinematic diffractions will build up to reconstruct the projected protein image on the detector. In our imaging method, we do not use the de-scan coil at downstream unlike conventional PED.

Higher tilt angle will effectively reduce dynamical diffractions, while we must be careful not to overlap the neighboring lattice points in higher Laue zones. The optimum tilt angle depends on the wavelength (electron beam energy) and the targeted highest spatial resolution. Figure 8 shows an example of the optimum condition, where Ewald sphere crosses the lattice point at target resolution point at right edge, in this case 1Å, i.e., $h = 10\,\mathrm{nm}/1\text{Å} = 100$. The radius of Ewald sphere is the vector length of the incident electron beam; $k_0 = 1/0.025 = 40\ \text{Å}^{-1}$. Therefore, the optimum title angle becomes $\varphi = 0.5\text{Å}^{-1}/k_0 = 12.5\,\mathrm{mrad} = 0.75°$. We must note that it crosses higher order lattice planes at opposite side, i.e., HOLZ: Higher Order Laue Zone, which causes unwanted noises thus we need to remove.

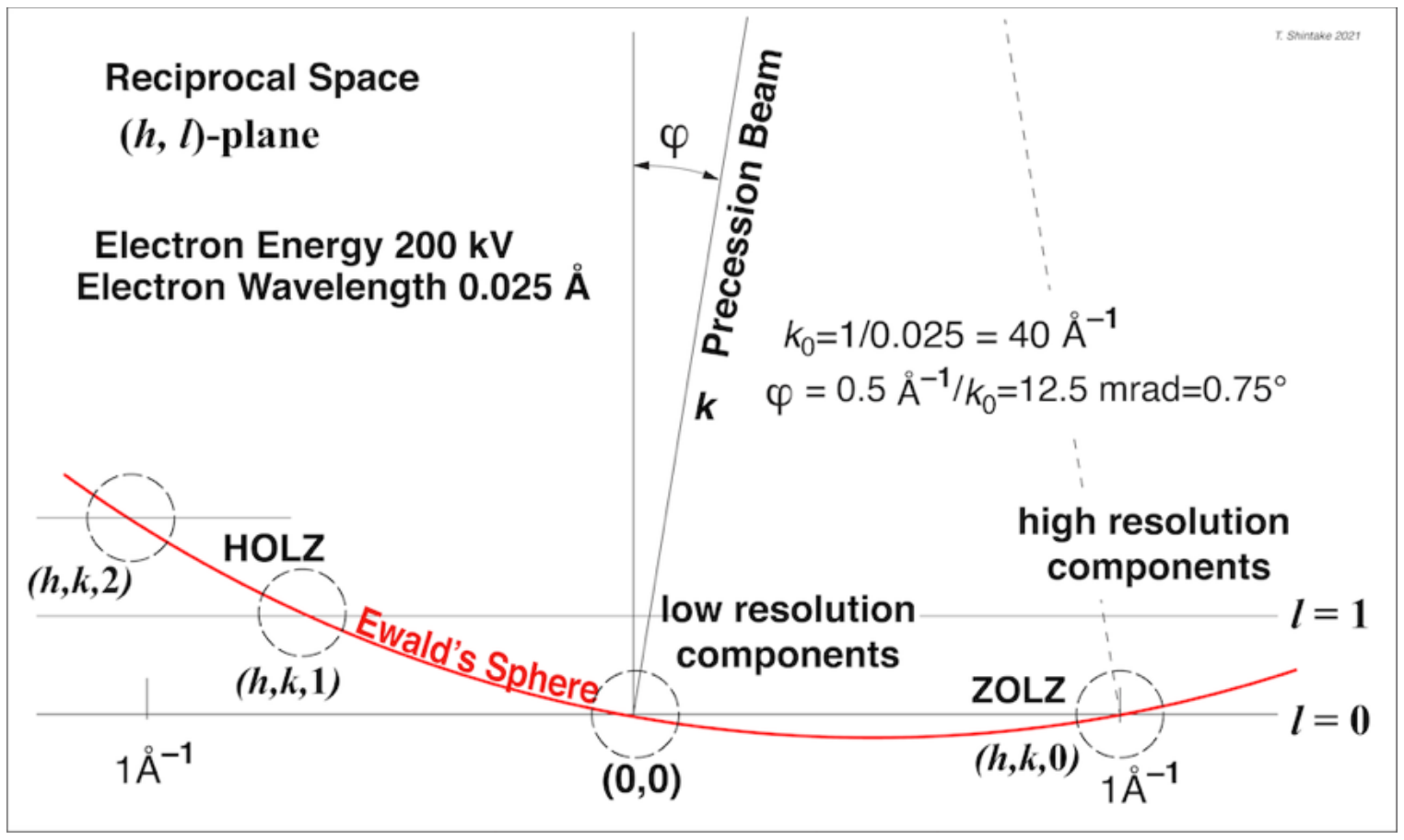

Fig. 8. Optimum precession angle for 200 keV beam and the target resolution of 1 Å.

Fig. 9 shows top view (*h,k*)-plane of the reciprocal lattice. Ewald sphere crosses the ZOLZ: Zero Order Laue Zone, $l = 0$) on a circle and lattices in a doughnut zone will diffract as the kinematic Bragg diffraction. Width of doughnut is determined by depth of the lattice point, i.e., $\Delta_l = 1/n_z$, while practical number does not affect on image quality because we precess the beam. The doughnut zone will rotate around the center on the reciprocal space, thus all of the lattice points within resolution rim of 1Å will diffract.

The long crescent zone at the left hand side is a part of HOLZ, which contains lattice points (*h,k,*1), represent longitudinal variation of protein form factor. Ewald sphere also hits (*h,k,*2) near to the resolution rim of 1Å$^{-1}$, whose interference fringe with the central beam will overlap with the fringe due to ZOLZ (*h,k,*0) at right edge at 1Å$^{-1}$. In principle, HOLZs and ZOLZ are part of the same crystal lattice, and they should contribute to reconstruct the real image constructively. However, as shown in Fig.10, HOLZs run through near edge of the objective lens, thus overfocused due to the spherical aberration, and cause frequency shift, resulting in strong beating with ZOLZ. This might be the source of Moiré patterns frequently observed in previously performed experiments [3]. Assume the right edge of ZOLZ: $p_1 = 1\text{Å}$, and left edge of HOLZ: $p_2 = 0.98 \times 1\text{Å}$. The pitch of Moiré pattern becomes

$$P = \frac{p_1 \cdot p_2}{p_1 - p_2} = 5\,\text{nm} \tag{15}$$

Randomly oriented crystals also cause this problem through the spherical aberration in the objective lens.

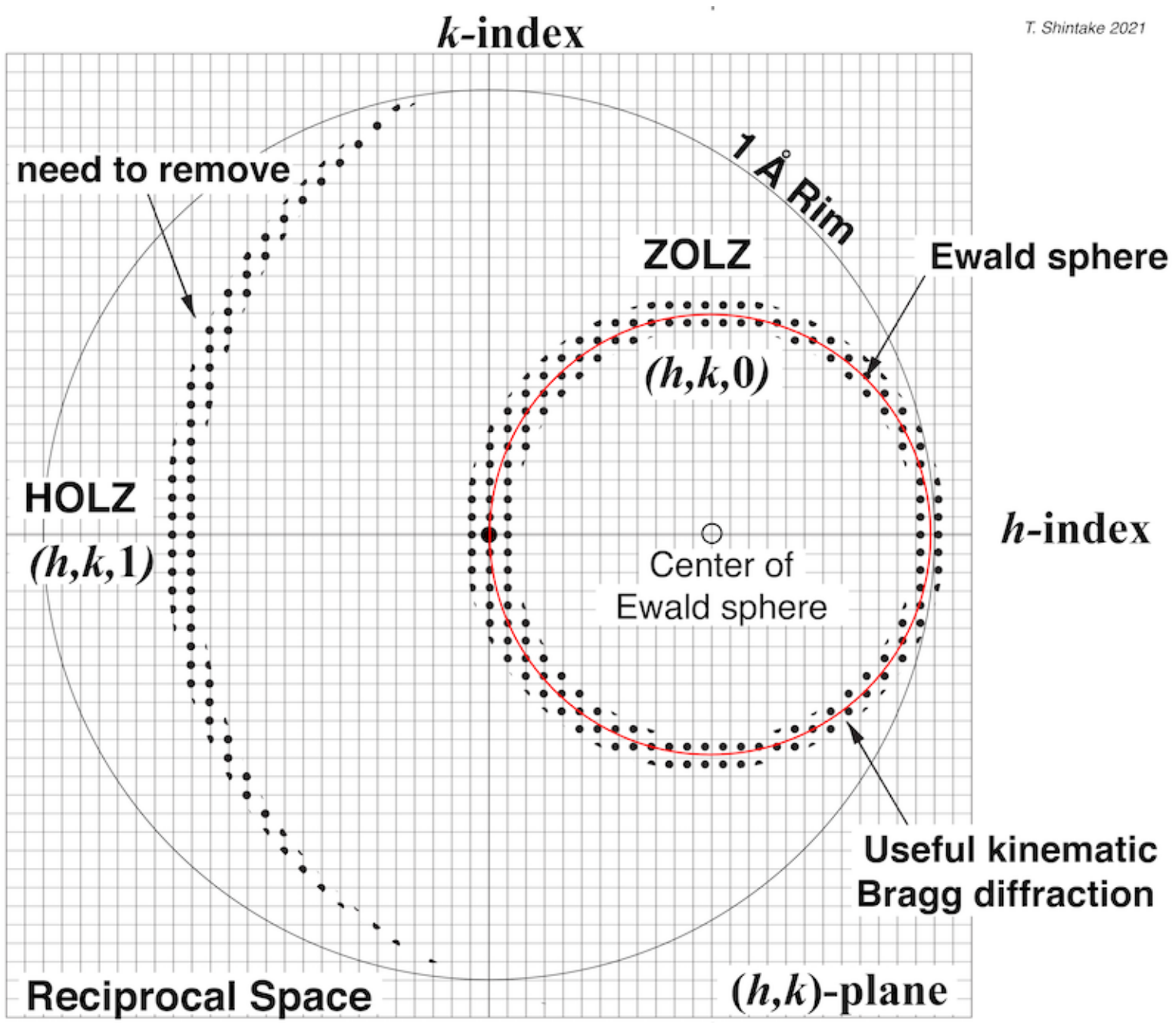

Fig. 9. Top view (*h,k*)-plane of the reciprocal lattice. Ewald sphere crosses the ZOLZ (zero order Laue zone, $l = 0$) on a circle and lattices in a doughnut zone will diffract as the kinematic Bragg diffraction.

Fortunately, we may remove those unwanted Moiré patterns as follows. Fig. 10 shows the wave vectors in the real space, where HOLZs take higher kick angle and thus we can remove them by means of an objective aperture at the focal plane. The central beam and the kinematic diffractions hit the doughnut zone on the objective lens and rotate around the optical axis. Therefore, if we place an annular objective aperture as shown in Fig. 10 at the focal plane, we may remove HOLZs.

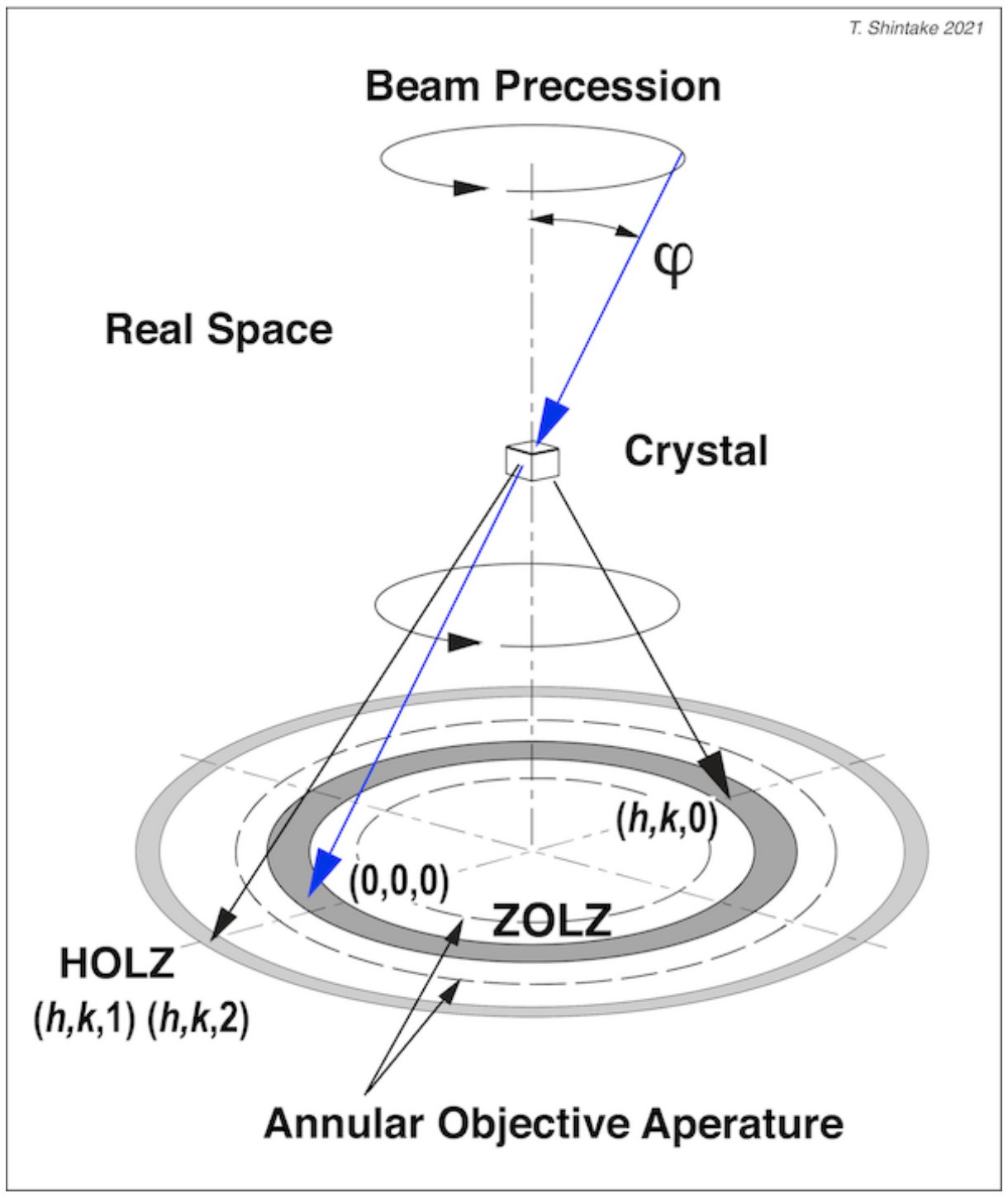

Fi. 10. The wave vectors in real space under precession electron diffraction.

In X-ray crystallography, there are commonly observed two concentric dark circles, corresponding to two consecutive diffraction orders of randomly oriented ice microcrystals associated with some defect of the cryo-protector or some humidity of the cold nitrogen used to cool down the sample. In our TEM measurement, these crystals diffract electrons into wide area and create noise background on the projected protein image. The annular aperture also eliminates those noise and increases contrast. The annular aperture is similar to that in the annular dark-field transmission electron microscopy (ADF-TEM). F. Leroux *et al*. [16] has successfully applied ADF-TEM to enhance contrast on imaging soft material by blocking central beam, and S. Bals *et al*. [17] used successfully in materials science where the annular aperture was applied to avoid diffraction contrast when recording a tilt series for tomographic reconstruction.

However, in our method, unlike conventional ADF-TEM, the central beam is not blocked and passes through the annular aperture and provides reference wave to the Bragg diffractions. It is bright field imaging. Our annular objective aperture works as a band-pass filter, which select only the kinematic Brass diffractions together with the central beam.

Figure 11 shows practical design of the annular objective aperture in our method. The diameter must match with the circular path of central beam under precession: $D = 2\varphi f$. It depends on practical microscope parameter; the diameter is roughly $30 \sim 100\,\mu\mathrm{m}$, which is easy to fabricate using FIB: focused ion beam tool. The width should match to the doughnut area $w \sim 2\varphi f / n_z$ in Fig. 9, while it should be wide enough to tolerate alignment errors between the precession beam to the aperture. The three bridges will generate Fresnel diffractions, but its shadow will disappear from the projection image because it is located at the focal plane (in the conventional optical camera, the shutter and iris are located at the focal plane because of this reason). To use this aperture, we need experimental verification and optimizations.

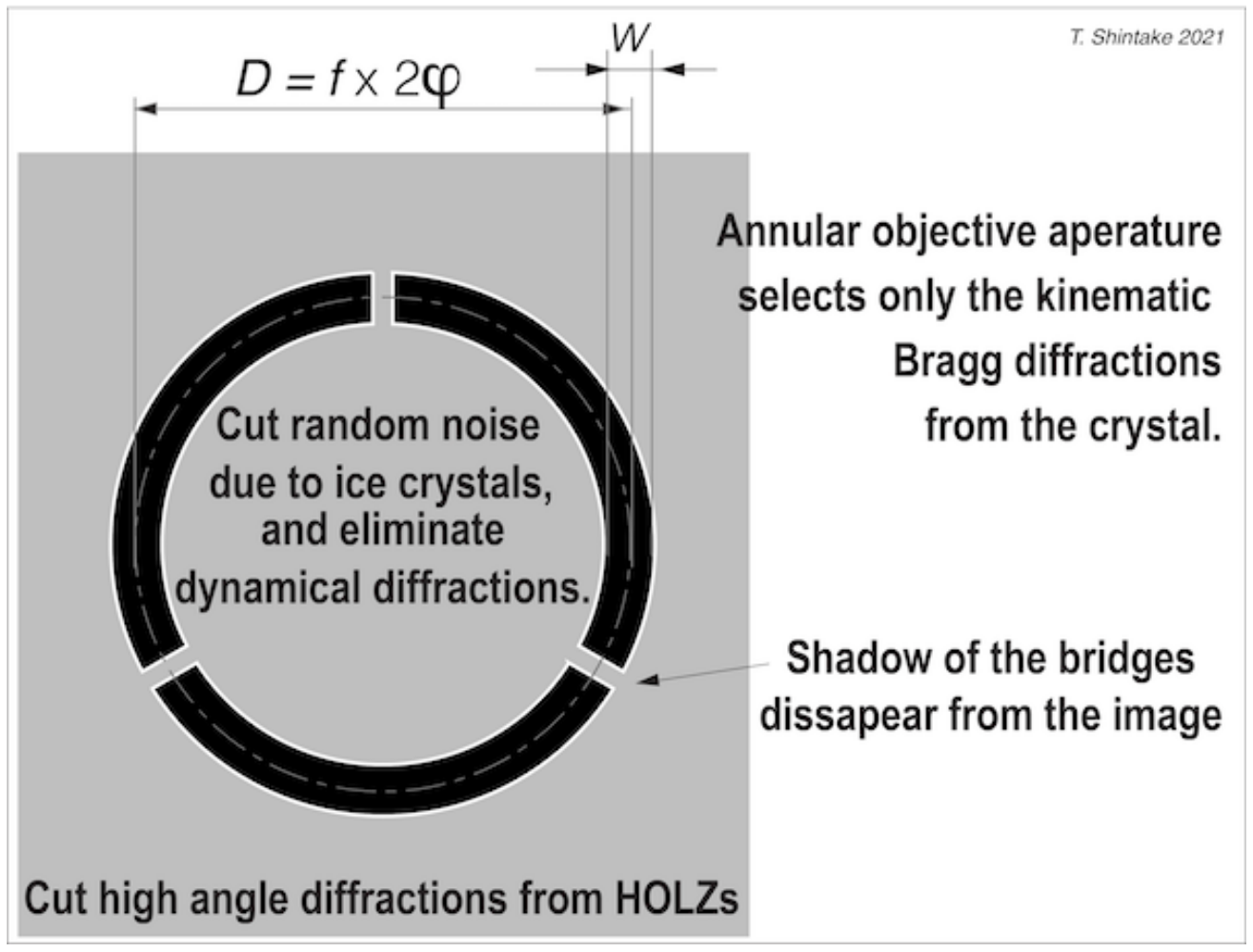

Fig. 11. Annular objective aperture for the precession electron microscopy.

Figure 12 shows conceptual illustration the conventional TEM image of microcrystal, and improved image with the beam precession and the objective aperture. Character “A” represents biomolecule in one unit cell; size of 10 nm x 10 nm. In the conventional method, Moiré patterns were frequency observed, shown by strong slanting lines. The dynamical diffraction causes reverse contrast, “A” is darker than surrounding area filled with exiting beams due to dynamical effect.

In the improved image, Moiré patterns and dynamical diffractions are eliminated, and “A” has positive contrast. From the quantum mechanical treatment, we know electrons are scattered by molecule, which contribute to reconstruct image at the same phase, thus, it results positive contrast. There are 10 x 10 = 100 unit cells in this illustration, and thus simply averaging them we should have roughly ten times of S/N improvement.

According to experimental data in Micro-EDs, protein crystals of 500 nm cubic keep their detail structure up to 10 e/ Å$^2$ electron dose. Using 2D electron detector of 4k x 4k pixel format, looking into 100 nm x 100 nm area of crystal, with corresponding pixel size 0.25 Å/pixel (twice of Nyquist frequency at 1 Å), we should have 1 x $10^7$ electrons in total, and 0.6 e/pixel. By averaging over 100 unit cells images, effective number of electron dose on single molecule becomes 1000 e/$Å^2$, which is quite high.

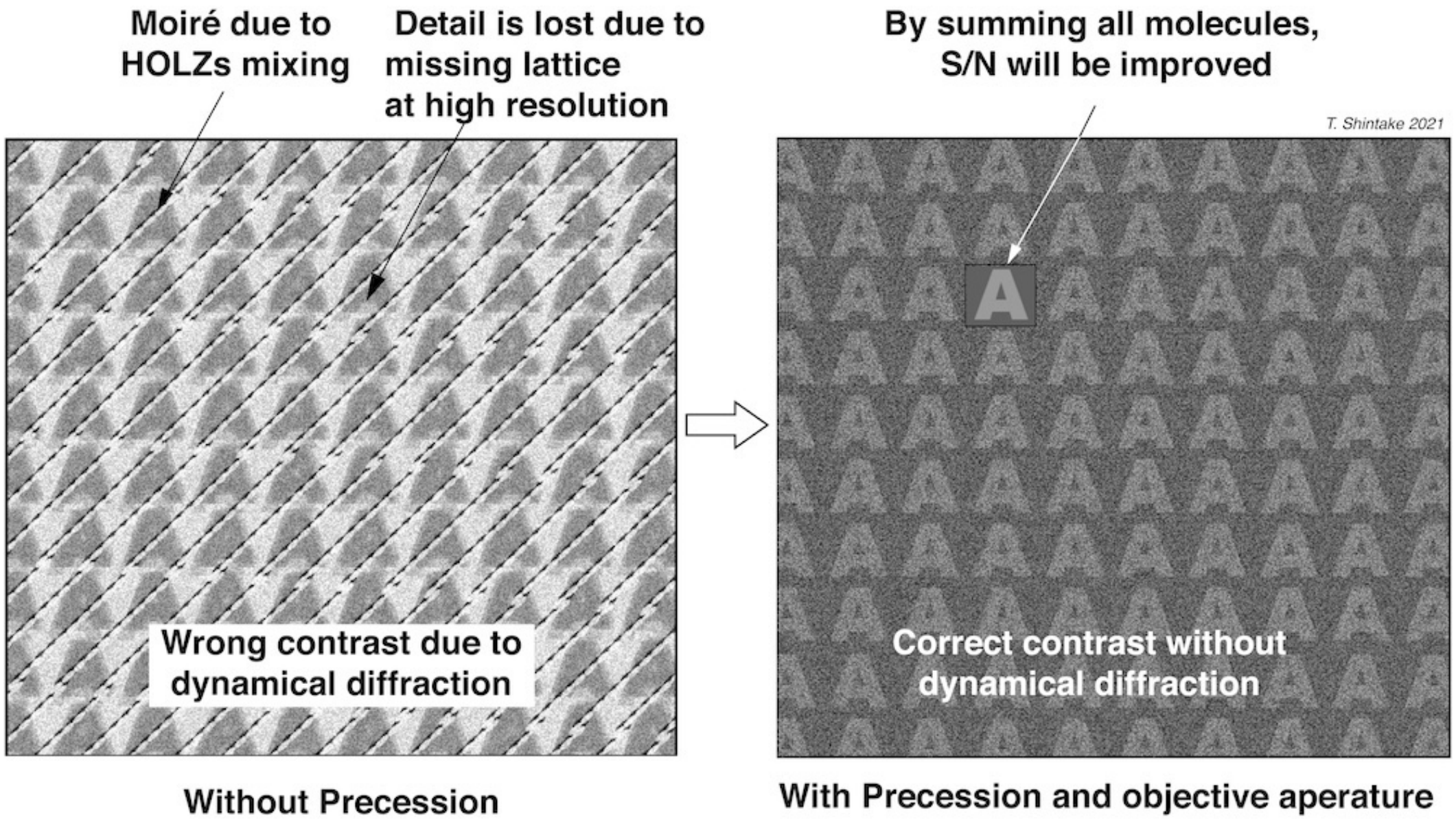

Fig.12. Conceptual illustration of the conventional TEM image of microcrystal, and improved image with precession beam and the objective aperture.

## VII. Spherical, Chromatic Aberrations and Coherency

### Spherical Aberration

The magnetic lens inherently has the spherical aberration problem, which limits available spatial resolution in common TEM. The beam deflecting angle through the object lens is a function of radial location of the beam; $dr/dz = -r/f - C_s\alpha^3/f$, where $C_s$ represents the spherical aberration. The central beam and high angle diffractions will have different focusing length, resulting in blurring image.

As we discussed in the previous section, we use the annular disk aperture right after the objective lens, which select only the kinematic Bragg diffraction and the central beam. All of them pass through the objective lens at the same radius: $r = f\alpha$, and receive the same deflecting angle, as a result, the reconstructed image will not be deteriorated by the spherical aberration. The effect of the spherical aberration cancels in this imaging method.

Figure 13 shows the detail beam path. In the precession electron microscopy, the incident beam is tilted at angle $\alpha$, which is reflected to angle $\alpha$ at the vertical lattice by Bragg diffraction, where the crystal lattice acts as vertical mirror because we align the crystal to the zone axis. Therefore, the central beam and Bragg diffraction go through the objective lens at two opposite points from the center $r = F_1\alpha$. In our beam expander system, two beams should be in parallel to the axis after the objective lens. However, due to the spherical aberration of the objective lens, the beam will be slightly over-focused by $C_s\alpha^3/F_1$, which causes spread at the image plane $2MC_s\alpha^3$. The crossing angle of two beams should be larger at the image plane:

$$\beta = \frac{\alpha F_1 - (F_1 + F_2)C_s\alpha^3/F_1}{F_2} + C_s\alpha^3/F_1$$
$$\approx \alpha/M - C_s\alpha^3/F_1 + C_s\alpha^3/F_1 = \alpha/M \qquad (16)$$

where we assumed $F_1 \ll F_2$. Two terms related to the spherical aberration are canceled. This is due to that the objective lens is located very close to the focal point of the projector lens (this is commonly satisfied in the large magnification system). Therefore, the crossing angle of two beams is not affected by the spherical aberration of the objective lens. We have to note that ray traces starting from center of the crystal causes longitudinal shift and spread on image plane due to aberration. However, the crossing angle of two traces stays the same, not influenced by the aberration. From Eq. (8), we find the pitch of the interference fringe as follows.

$$p' = \frac{\lambda}{sin(2\beta)} \sim M\,\frac{\lambda}{2\alpha} = M \cdot p \qquad (17)$$

The spherical aberration disappears. We have to note that the path difference between two traces is zero, and thus interference fringe position is unchanged by the spherical aberration. As shown in Fig. 5, the interference fringe pattern is longitudinally elongated, thus shift of crossing point does not affect to the fringe position on the detector. This is the reason why the image is insensitive to the focus also.

We may conclude the spatial resolution of the accumulated image may not be limited by the spherical aberration of the objective lens, in this ideal condition. This is surprising conclusion, and totally different from the common imaging method in the transmission electron microscopes. The resolution will be limited by effective electron source size (transverse coherency), the vibration and the drift of sample. Experimental investigation must be performed to verify the realistic limitations.

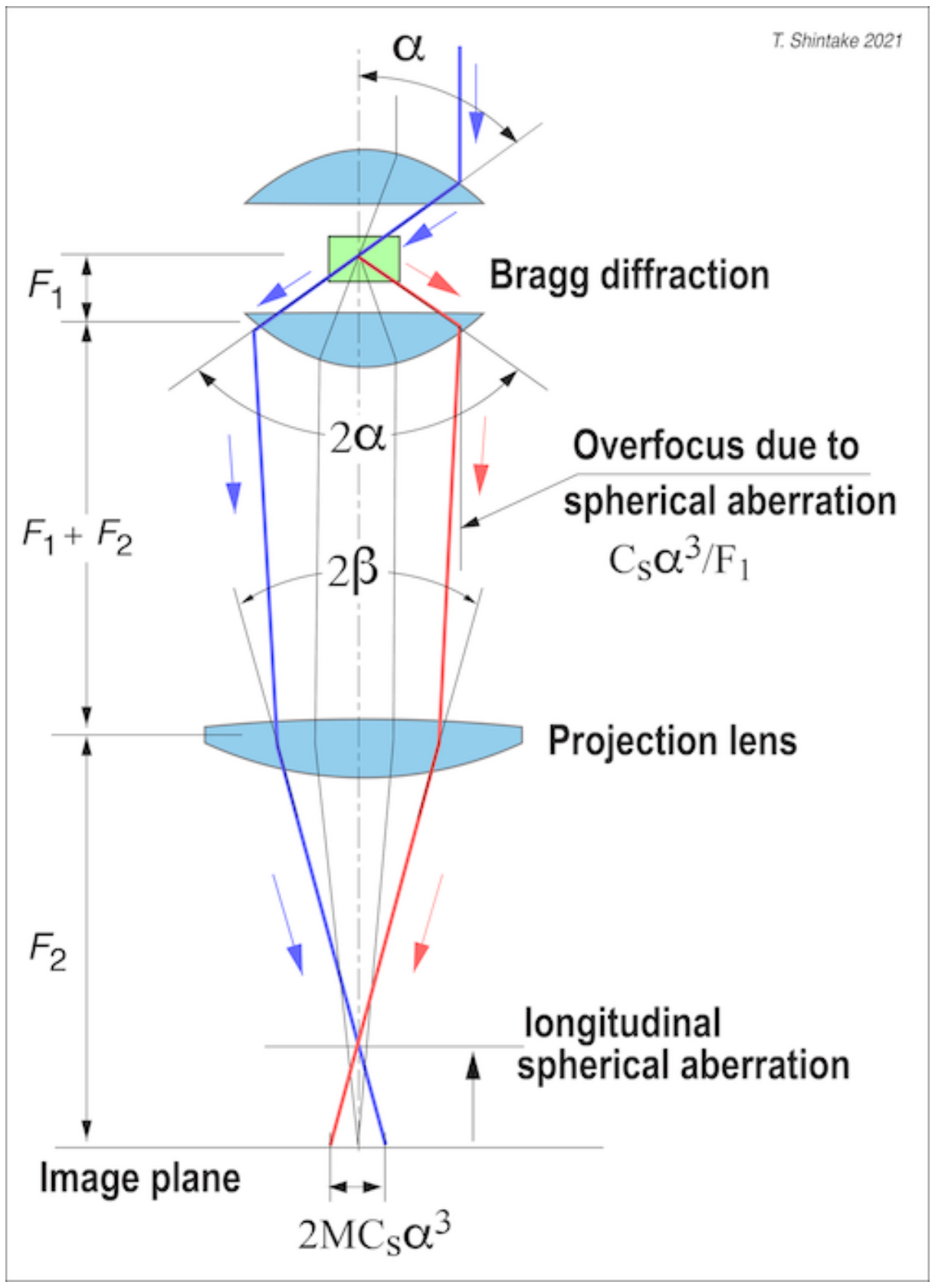

Fig. 13. Spherical aberration effect.

## Chromatic Aberration

The energy spread dilutes image quality through chromatic aberration of objective lens in common TEM. However, if the crystal and the lens system are well aligned, the energy spread does not directly deteriorate available resolution, in our method. Energy deviation causes focusing position shift, just same as longitudinal spherical aberration does, while it does not deteriorate image quality, since this method is insensitive to the defocus.

## Coherency and others

As we learned from Tonomura experiment [13], the interference effect is "single electron event". In Fig. 13, two electrons run through left and right arms, and meet at the detector, case interference effect. This is wrong interpretation. Individual electron is fully coherent, i.e., longitudinally and transversely. Inside the crystal, a part of the electron wave is reflected to the right, and the transmission wave goes to the left. They meet at the detector and create event, whose probability is governed by $|\Psi|^2$, represents interference pattern (see Appendix A3).

Individual electrons start from emitter surface, randomly in time, in position (source size) and in energy (energy spread). The effective source size can be made much smaller than actual emitting surface area by means of small irises in the condenser system, while available electron current becomes low. The cold field emission gun (C-FEG) has high brightness ~2x10$^9$ A/cm$^2$sr, and thus provide enough dose, suitable for molecule imaging at angstrom resolution. Requirement on source parameter should be the same level as common TEM for cryo-microscopy.

When an electron collides to atoms inside the crystal, the electron wave collapse on the interacting atom, deposits a part of the kinetic energy, and again starts different electron wave, which does not interfere with original wave. This is the in-elastic collisions, do not contribute to the molecule imaging, but causes random noise around the initial beam. The annular objective aperture partly reduces this noise, but it still increases background. We need to introduce digital image processing to effectively reduce those noise based on FFT: Fast Fourier Transform.

The thermal vibration of atoms causes blurring on molecule image. Vibration amplitude highly depends on flexibility of protein inside the crystal. For well-ordered structure, blurring radius is less than 0.6 Å [12].

## VIII. Electron Dose and 3D Reconstruction

For small molecule such as chemicals or drugs (< 1 kD for example), a few projection images will be enough to identify the structure, and thus useful for rapid screening, including chirality. Those materials are more radiation resistant than the protein crystals, and thus we will be able to obtain high resolution image with high S/N (signal-to-noise ratio).

For protein crystals, according to the experiences in Micro-ED, maximum dose 10 e/Å$^2$ is allowable with 200 keV beam. For example, we divide this dose in three tasks.

(1) Find a target crystal in low magnification TEM mode, refer to the next section.

(2) Take still image with precession beam at high magnification and high resolution.

(3) Turn crystal to the nearest zone-axis a few degrees apart, such as [001] to [1,0,20]. Take still image with precession beam. Due to the radiation damage, detail structure might be already lost, but it will be useful to obtain rough estimate of depth information.

By combining two images [001] and [1,0,20], we may create "stereo view", from which we determine relative location of the building-blocks-amino acids and small peptides or alpha-helix. For example, let's assume diameter of molecule is 100 Å, to identify the same alpha-helix (diameter of 5 Å) on the outer surface, the angle should be smaller than 5/100 = 0.05 rad = 3 degree. The angle between two axes [001] vs [1,0,20] is 1/20 = 0.05 rad = 3 degree, and thus we will be able to track the position shift of the same alpha-helix without missing and confirm its location, most importantly front or back sides.

We repeat the same processes on different crystals with different zone axes. By combining 13 unique zone axes (see Appendix 1), we will be able to determine 3D structure of large protein complex. Alternatively, we may collect multiple projection images at surrounding edges of a larger crystal, by rotating crystal to successive zone axes.

It will be also possible to combine with diffraction data from Micro-ED, where we may provide phase information by 2D FFT on projected images. This is the same idea as Henderson and Unwin [1, 2] used in 2D protein crystal analysis.

X-ray CT uses multiple projection image to reconstruct 3D structure. It will be possible to use the same working principle of Radon transform [20] for protein structure analysis. We have to note that limited number of projection images will induce unwanted traces in reconstructed structure. We need further R&D to combine Radon transform with constraint of localized density to the building blocks.

## IX. Practical Instrumentation

Figure 14 shows the schematic illustration of proposed precession electron microscopy. The electron beam runs from top to down.

(1) We use high resolution cryo-TEM, optimized for bio-molecule imaging. 200 kV with the cold field emission gun (C-FEG) will be the best choice. The vacuum system around the sample must be well designed to avoid microcrystal grow on the cryo-sample due to residual gas.

(2) Double tilt cryo-holder will be one of the key components. It should be motorized design. We must note that the drift and vibration will directly deteriorate the output image.

(3) We do not need Cs corrector. In this microscopy method, it is important to have cylindrically symmetric field in the optical system. The remanent fields of quadruple and sextapole fields inside the Cs corrector will be harmful.

(4) The annular objective aperture is the most important component for clean image acquisition. Quick in-out capability will be required in the practical operation, accurate repeatability on position setting must be realized.

(5) For the beam precession, set of beam deflector coils may be digitally controlled with highly accurate AD conversion. Special control software will be required to tune the beam axis through the objective lens, the annular objective aperture and the projection system.

(6) Direct electron 2D detector will be the best choice for low dose imaging, which should have 4k x 4k pixel format or higher. For better imaging outcome, we need to capture the projection image as wide as possible in one shot, and thus we gain higher S/N after averaging on multiple unit cells. At the same size, the magnification should be high enough to secure the Nyquist frequency higher than target spatial resolution. Therefore, larger pixel format is required.

(7) As experienced in cryo-electron microscopy, sample always shows drift due to the electron beam irradiation and ice expansion or the electric charging effect. Fast frame camera has been used to track the motion and correct. We may apply the same method.

(8) The alignment of the crystal axis is technical challenge. The diffraction pattern has been used to find crystal axis traditionally. However, switching between diffraction mode and imaging mode will take time to settle the magnetic field of the lens. And also, the central beam has risk of damage on the direct electron detector.

Presumably, we may find target crystal at zone axis as follows. By running upstream optics as beam precession, using TEM mode of wide view thus extremely low current density on sample, taking wide area image (lower magnification) through annular objective aperture on live mode, scan rotation angle of crystals using double tilt sample holder, where some of crystals light up at zone axis (assume we have many microcrystals on the grid). Outside the zone axis condition, the shadow of crystal should be dark (there is not many electrons passing annular aperture). Zoom into target crystal, focused area illumination (keep parallel beam by tuning condenser lenses) thus higher current density, take projection image within damage threshold. We do not need to use diffraction mode, thus there is no risk of damage on the expensive direct electron camera.

After taking images from multiple microcrystals, we classify the images according to axes and orientation, and reconstruct 3D structure same as single particle cryo-electron microscopy. The stereo image discussed previous section may help this process.

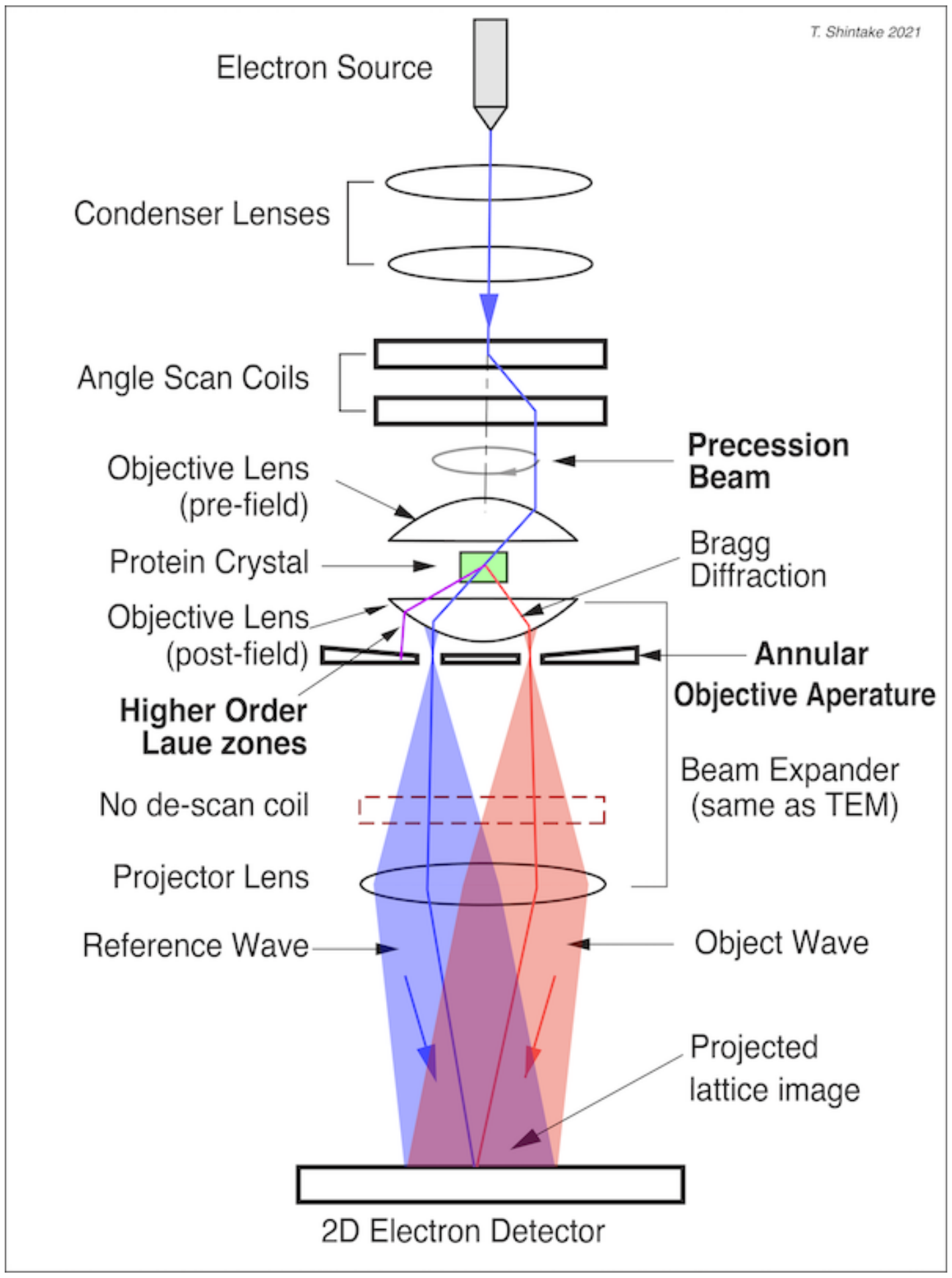

Fig. 14. Schematic illustration of the precession electron microscopy for the protein crystal.

## X. Conclusions

In this paper, the author proposed Bragg diffraction imaging and the precession electron microscopy, suitable for projection imaging on the protein microcrystals. The features of this method are;

(1) Using PED: Precession Beam Diffraction to reduce dynamical diffraction.

(2) Using annular objective aperture to select only the kinematic Bragg diffractions, and remove various noises. Unlike conventional ADF-TEM, this method is the bright-field imaging, because the central beam passes through the aperture, and provides the reference wave.

(3) Image is insensitive to defocus, and the spatial resolution is not limited by the spherical and the chromatic aberrations of the objective lens.

As experienced in Micro-ED community, experimental results are highly dependent on quality of the protein crystals, and detailed procedure of the data taking (rotation angle step, dose of electrons, etc). They are related to technical details on the electron microscope and its control software. Therefore, the author encourages collaboration between scientists from the protein crystallography and the cryo-electron microscopy, together with the electron microscope industry to realize the proposed imaging method. Optimizing the protein crystallization is also an important key issue. Mosaic and disorders will be directly observed, which may help to screen the sample and improve crystallization method.

## Acknowledgement

The author thanks to Dr. Masao Yamashita and Dr. Ryusuke Kuwahara for their encouragement and discussions on the nature of protein crystals and issues in the cryo-electron microscopy. The author acknowledges to Dr. Cathal Cassidy for his useful discussions on technical details in the transmission electron microscopy.

# Appendix

## A1: Zone Axis [21]

Figure A1 shows standard cubic stereographic projection of dominant zone axes. There are $3 \times 3 \times 3 = 27$ combinations, while [000] does not have unit vector, so that number of possible vectors is 26. We know the reverse index should have the same projection image; $I(\overline{uvw}) = I(uvw)$. All points back side of the paper will be same as that in front side of the paper. Eliminating same zone axes, such as, $[\bar{1}00] = [100]$, we have 13 unique zone axes, on which we should take images. We may take image at any higher order axes, such as, [513], while Bragg diffraction pattern may be sparse and non-symmetric, and thus reconstructed projected image becomes not clear. This is due to overlapping with surrounding molecules, in real space.

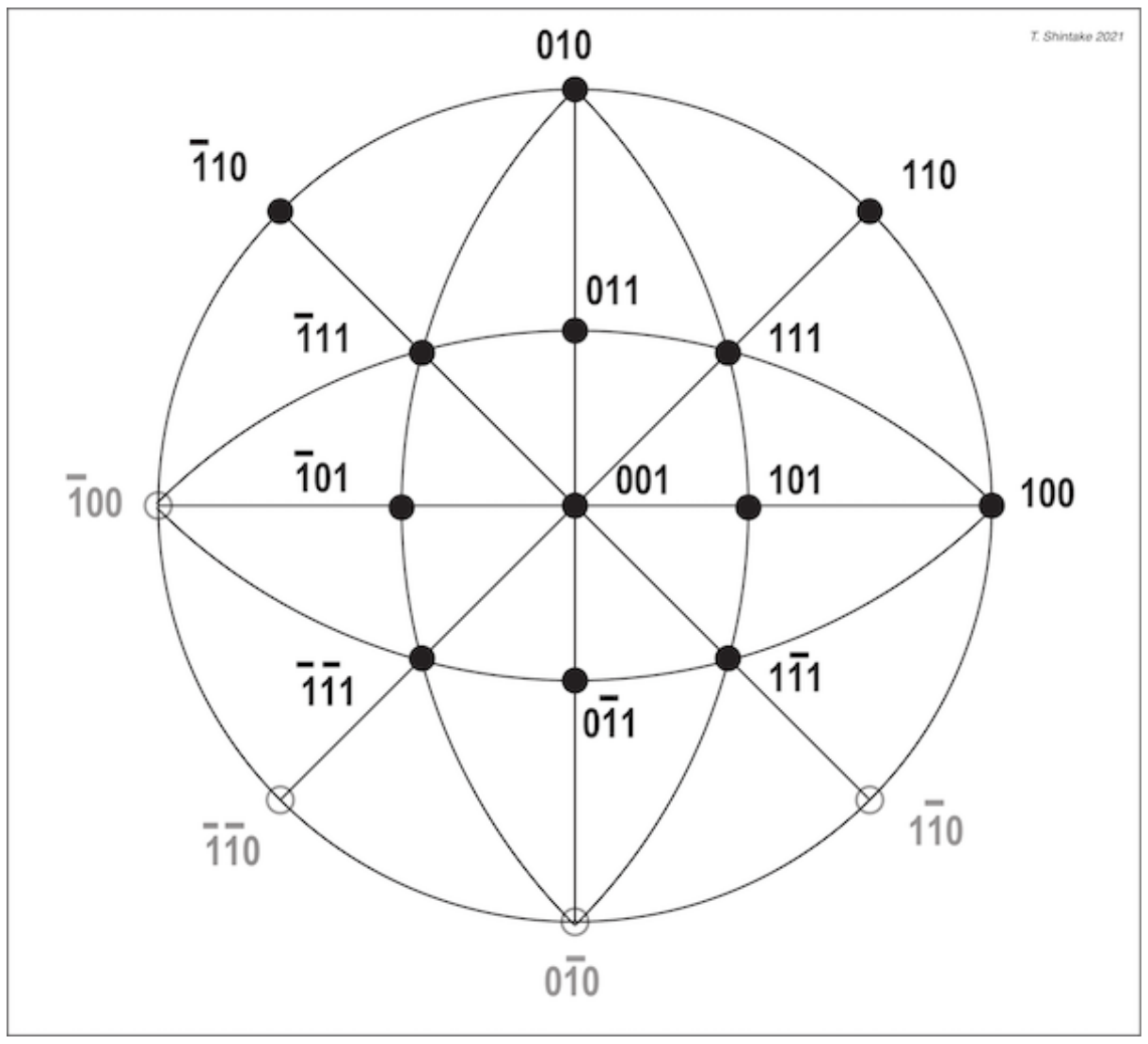

Fig. A1. Standard cubic stereographic projection of dominant zone axes on half hemisphere. Unique 13 zone axes are shown black dots with bold font label.

## A2: The Beam Expander

The beam expander is frequently used in visible-light optics to expand (or reduce) beam size to match with required magnifications. The diagram of Fig. A2 shows Keplerian type beam expander consists of two convex lenses. There is another type: Galilean beam expander, which uses concave lens to expands beam, but it cannot be realized in the electron beam optics, thus we do not treat here.

Fig. A2 (a) is the beam expander. By combining two lenses with different focus length and we arrange the total travel length is equal to sum of these focusing lengths (common focus). In the incoming parallel beam is expanded its size, and output becomes parallel beam. All light rays go through the common focus point and spread; thus, image is reversed. The magnification factor is $M = F_2 / F_1$ .

Fig. A2(b) is common microscope. By using the same optical configuration, a point source $O$ is located at distance $F_1$ upstream of the lens, the parallel beam is created and focused into spot $O$' at distance $F_2$ from the lens. If the half opening angle of emission beam is $\alpha$, the focusing angle becomes $\alpha /M$. If there are two points at the object; $O_1$ and $O_2$ with transverse separation $\delta$, they are magnified and projected to image plane; $O_1$' and $O_2$' with separation $M\delta$. If we use today's electron microscope, we may achieve $M = 1 \sim 50 \times 10^6$. Fig. A2(b) is also identical to "infinity corrected optics" of the objective lens in optical microscope. Between two lenses, the beam becomes parallel optical path, thus we may choose distance between two lenses freely, and allow us to implement additional optical components such as polarizer or illuminator.

Fig. A2(c) is more general case, the parallel beam passing point $O$ with angle $\alpha$, the output beam becomes parallel beam of its size $M$ times magnified and angle of $\alpha /M$.

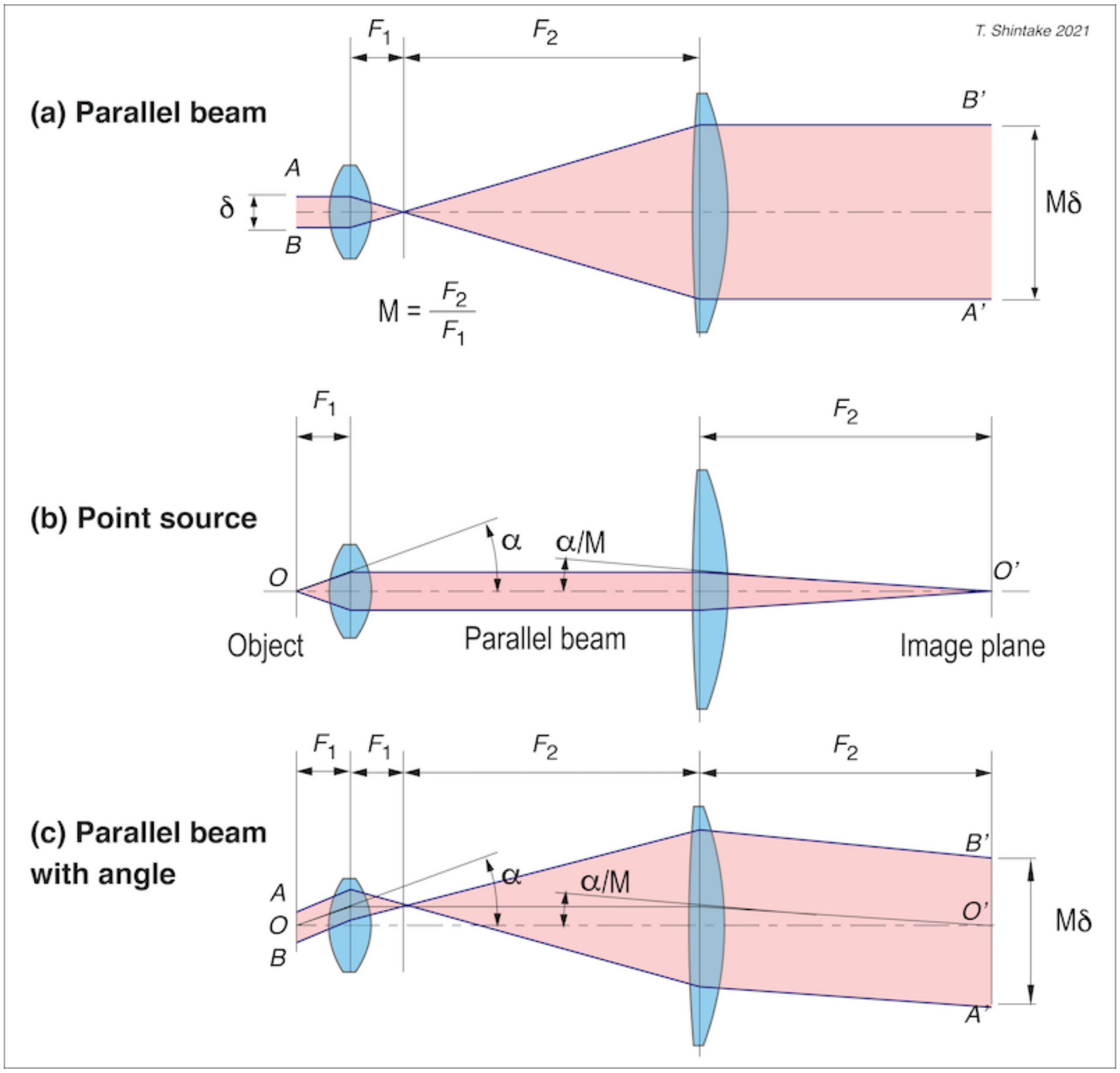

Fig. A2. Three optics based on identical two convex lenses.

**Appendix A3: Two-beam interference, Young's experiment.**

Figure A3 shows details of two beam interference. When two plane waves incident on a detector, we observe periodic intensity modulation. The wave may be the light wave, the electron wave or the neutron wave. At the center of the bright zone, two waves incident to the detector at the same phase. If the phase of Wave-1 changes $\pm\Delta\phi$, the bright zone shifts accordingly. Therefore, the phase difference of incident wave is recorded as fringe location. In crystal imaging, information of the lattice location is transferred through Bragg diffraction, and recorded on the electron detector.

We have to note that the amplitude has the standing wave pattern, oscillating plus and minus. Fourier transform of the amplitude shows two spots, corresponds to forward and backward waves. Fourier transform of the power, time average of amplitude square shows three components. Two side bands should contain the phase. However, the phase is missing in the recorded diffraction pattern, because only the intensity is recorded. This is the phase problem.

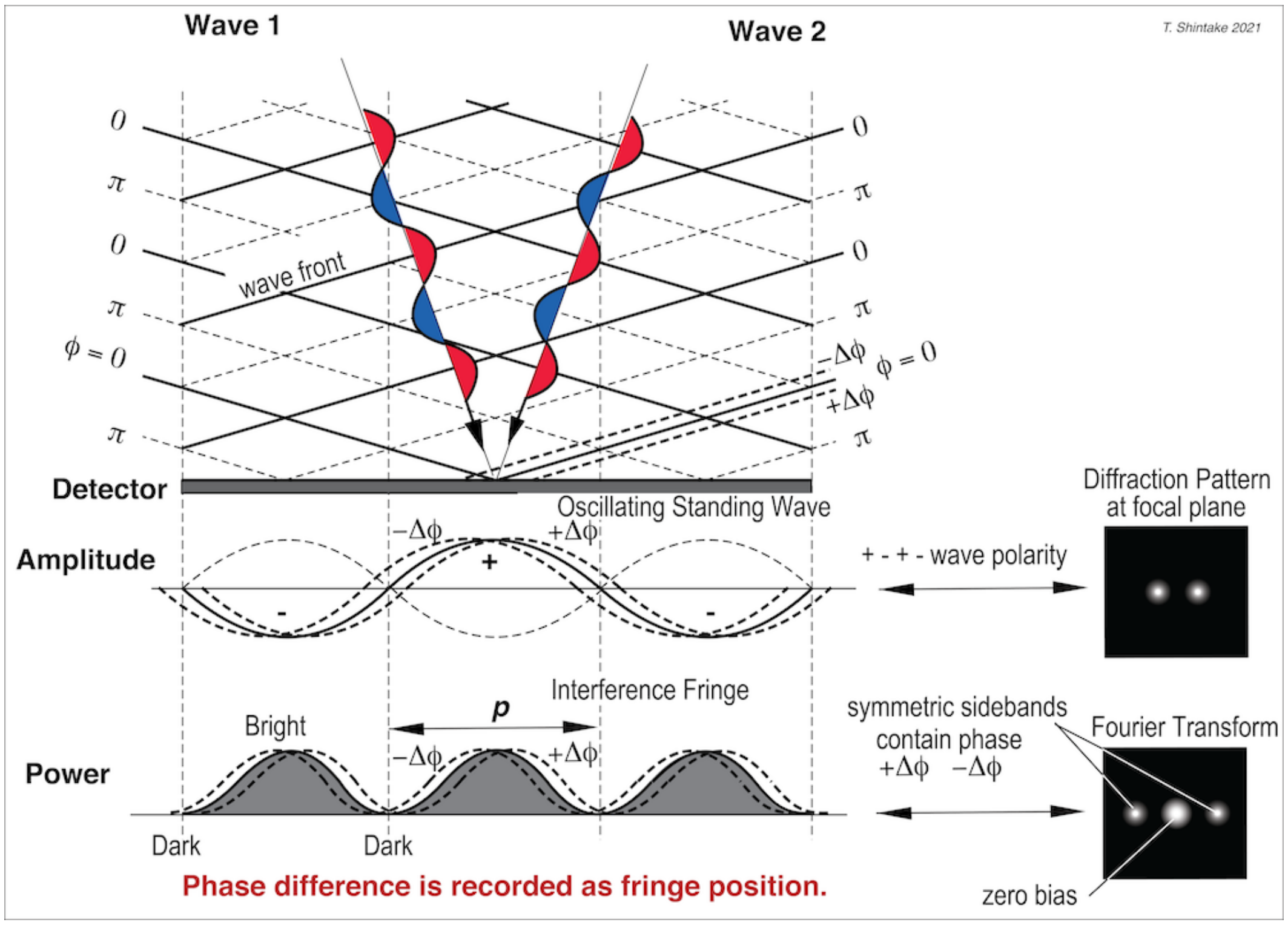

Fig. A3. The two-beam interference. The phase difference between two incoming waves is recorded as transverse position of the interference fringe, i.e., the lattice location.